\documentclass[12pt,showpacs,fleqn,preprintnumbers,amsmath,amssymb]{article}
\usepackage{graphicx}
\usepackage{amssymb}
\usepackage{amsmath}
\makeatletter \oddsidemargin -0.1in \evensidemargin -.1in
\newcommand{\singlespacing}{\let\CS=\@currsize\renewcommand{\baselinestretch}{1.5}\tiny\CS}
\newcommand{\doublespacing}{\let\CS=\@currsize\renewcommand{\baselinestretch}{1.5}\tiny\CS}
\begin{document}
\begin{center}
\textbf{Effect of Landau damping on ion acoustic solitary waves in a multi-species collisionless unmagnetized plasma consisting of nonthermal and isothermal electrons}\\
Sandip Dalui and Anup Bandyopadhyay\\
\textit{\ Department of Mathematics, Jadavpur University, Kolkata - 700 032, India.}
\end{center}

\noindent \textbf{Abstract} $-$ A Korteweg-de Vries (KdV) equation including the effect of Landau damping is derived to study the propagation of weakly nonlinear and weakly dispersive ion acoustic waves in a collisionless unmagnetized plasma consisting of warm adiabatic ions and two different species of electrons at different temperatures. The hotter energetic electron species follows the nonthermal velocity distribution of Cairns et al. [Geophys. Res. Lett. \textbf{22}, 2709 (1995)] whereas the cooler electron species obeys the Boltzmann distribution. It is found that the coefficient of the nonlinear term of this KdV like evolution equation vanishes along different family of curves in different parameter planes. In this context, a modified KdV (MKdV) equation including the effect of Landau damping effectively describes the nonlinear behaviour of ion acoustic waves. It has also been observed that the coefficients of the nonlinear terms of the KdV and MKdV like evolution equations including the effect of Landau damping, are simultaneously equal to zero along a family of curves in the parameter plane. In this situation, we have derived a further modified KdV (FMKdV) equation including the effect of Landau damping to describe the nonlinear behaviour of ion acoustic waves. In fact, different modified KdV like evolution equations including the effect of Landau damping have been derived to describe the nonlinear behaviour of ion acoustic waves in different region of parameter space. The method of Ott \& Sudan [Phys. Fluids \textbf{12}, 2388 (1969)] has been applied to obtain the solitary wave solution of the evolution equation having the nonlinear term $(\phi^{(1)})^{r}\frac{\partial \phi^{(1)}}{\partial \xi}$,  where $\phi^{(1)}$ is the first order perturbed electrostatic potential and $r =1,2,3$. We have found that the amplitude of the solitary wave solution decreases with time for all $r =1,2,3$.

\begin{center}
\textbf{1. INTRODUCTION}
\end{center}

Two different species of electrons at different temperatures are common in laboratories [1-4] and also in space plasmas [5-6]. Plasmas consisting of two different species of electrons at different temperatures have been observed by various spacecraft missions, viz., Freja Satellite [7], FAST at the auroral region [8-12], Viking Satellite [13-14], S3-3 Satellite [6], GEOTAIL [15] and POLAR [12,16,17] missions in the magnetosphere. These satellite observations also indicate that the distribution functions of both the electron species are not necessarily isothermal for the existence of two - electron - temperature plasmas. In particular, the FAST [8-12] satellite and the Viking Satellite [13-14] observations of the electric field structures in the auroral zone along with the observations of electric field structures by the Freja Satellite [7] in the auroral zone of the upper ionosphere indicate the presence of a population of fast energetic electrons together with isothermal electrons. Among these two species of electrons, the cooler electron species follows Boltzmann-Maxwellian distribution whereas the hotter electron species can be described by considering Cairns [18] distributed nonthermal electrons. 

Several authors [19-28] investigated different linear and nonlinear properties of ion acoustic (IA) waves in a multi-species plasma consisting of one or two ion species and two different electron species at different temperatures. Islam et al. [24] investigated the nonlinear behaviour of IA waves in a collisionless magnetized plasma consisting of warm adiabatic ions and two different populations of electrons at different temperatures, where one species follows Boltzmann-Maxwellian distribution and the other one obeys the nonthermal Cairns [18] distribution. Recently, Dalui et al. [29] have investigated the modulational instability of IA waves in a collisionless unmagnetized plasma consisting of warm adiabatic ions, nonthermal and isothermal electrons. Later Dalui et al. [30] have considered the same problem in presence of external static uniform magnetic field. Dalui et al. [29] have rigorously considered the problem of validity of describing the electrons as two different species of particles which are following two different types of distribution functions with respect to the ion acoustic time scale. In the present paper, we have considered the same plasma system of Dalui et al. [29] to investigate the nonlinear behaviour of IA waves along with a correction due to the kinetic effects of two different species of electrons.     

Longitudinal electron plasma oscillations are damped during the propagation through a collisionless plasma. In particular, Vlasov [31] used the linearized Boltzmann equation to investigate the small amplitude steady state longitudinal electron plasma oscillations. Shortly afterwards, Landau [32] pointed out that these oscillations are damped. This damping of longitudinal electron plasma waves in a collisionless plasma is known as linear electron Landau damping. There are several interesting papers [33-42] on the existence and physical interpretation of linear electron Landau damping and experimentally this phenomenon was verified by Malmberg and Wharton [43]. Grishanova \& Azarenkova [44] studied electron Landau damping in toroidal plasma with Solov'ev equilibrium. For the first time, Ott and Sudan [45] investigated the effect of linear electron Landau damping on IA solitary waves in a collisionless plasma. Several authors investigated the effect of Landau damping on IA solitary waves in unmagnetized or magnetized plasmas theoretically [46-52] and experimentally [53]. In particular, Tajiri and Nishihara [47] investigated the effect of Landau damping on finite amplitude IA solitary waves in a collisionless unmagnetized electron-ion plasma consisting of cold ions and two distinct populations of isothermal electrons at different temperatures by considering a KdV like evolution equation including the effect of Landau damping. Bandyopadhyay and Das [48] derived a Korteweg-de Vries-Zakharov-Kuznetsov (KdV-ZK) and a modified KdV-ZK (MKdV-ZK) equations including the effect of electron Landau damping to investigate the nonlinear behaviour of IA waves in a magnetized plasma consisting of warm adiabatic ions and nonthermal electrons. Again, Bandyopadhyay and Das [49] investigated the effect of electron Landau damping on kinetic Alfv$\acute{\mbox{e}}$n and IA solitary waves in a magnetized nonthermal plasma with warm ions. Recently, Ghai et al. [54] investigated the dust acoustic solitary and shock structures under the influence of Landau damping in a dusty plasma containing two different temperature ion species.

In the present paper, we have considered a collisionless unmagnetized electron-ion plasma consisting of two distinct populations of electrons at different temperatures, one of which obeys the Maxwellian velocity distribution, and the other electron species follows the nonthermal distribution of Cairns et al. [18]. To investigate the effect of linear electron Landau damping on IA solitary waves, we have considered coupled Vlasov - Poisson model for two different electron species along with the fluid model for ions. So, in the present plasma system, the kinetic effects of two different species of electrons at different temperatures have been investigated on IA solitary structures with special emphasis on the following cases:

\noindent \textbf{Case - 1:} Using the reductive perturbation method, an evolution equation has been derived which describes the nonlinear behaviour of IA waves along with a correction due to the kinetic effects of two different species of electrons. This evolution equation reduces to a well known Korteweg-de Vries (KdV) equation if electron to ion mass ratio is neglected.

\noindent \textbf{Case - 2:} It is found that a factor ($B_{1}$) of the coefficient of the nonlinear term of the evolution equation derived in Case - 1 vanishes along different family of curves in different parameter planes. In this situation, i.e., when $B_{1}=0$, a modified evolution equation including the effect of electron Landau damping efficiently describes the nonlinear behaviour of IA waves and this modified evolution equation becomes a modified KdV (MKdV) equation having the nonlinear term $\displaystyle (\phi^{(1)})^{2}\frac{\partial \phi^{(1)}}{\partial \xi}$ if electron to ion mass ratio is neglected, where $\phi^{(1)}$ is the perturbed electrostatic potential and $\xi$ is the stretched space variable.

\noindent \textbf{Case - 3:} It has been observed that a factor ($B_{2}$) of the coefficient of the nonlinear term of the evolution equation derived in Case - 2 vanishes along a family of curves in the parameter plane. In this context, a further modified evolution equation including the effect of electron Landau damping can describe the nonlinear behaviour of IA waves when the conditions $B_{1}=0$ and $B_{2}=0$ hold simultaneously and this further modified evolution equation reduces to a further modified KdV (FMKdV) equation having nonlinear term $\displaystyle (\phi^{(1)})^{3}\frac{\partial \phi^{(1)}}{\partial \xi}$ if electron to ion mass ratio is neglected. For the first time, we have derived a FMKdV equation having nonlinear term $\displaystyle (\phi^{(1)})^{3}\frac{\partial \phi^{(1)}}{\partial \xi}$ including the effect of electron Landau damping. 

\noindent \textbf{Case - 4:} Using the multiple time scale analysis of Ott and Sudan [45], we have developed a general method to find the solitary wave solution of the evolution equation having nonlinear term $\displaystyle (\phi^{(1)})^{r}\frac{\partial \phi^{(1)}}{\partial \xi}$ including the effect of electron Landau damping.

\noindent \textbf{Case - 5:} The amplitudes of the solitary wave solutions of the different evolution equations including the effect of electron Landau damping have been investigated for $r =1,2,3$ and it is found that the amplitude of the solitary wave solution decreases with time for all $r =1,2,3$.  

\begin{center}
\textbf{2. BASIC EQUATIONS}
\end{center}

In the present paper, we have considered a collisionless unmagnetized plasma consisting of warm adiabatic ions and two different species of electrons at different temperatures. The hotter electron species is nonthermally [18] and the cooler one is isothermally distributed. To describe the nonlinear behaviour of IA waves propagating along $x-$axis including the correction due to the kinetic effects of electrons, we take the Vlasov - Poisson model for two different electron species and the fluid model for ions. The Vlasov - Poisson model for two different electron species at different temperatures can be written by the following equations:
\begin{equation}\label{m_f_ce}
 \sqrt{\frac{m_{e}}{m}} \frac{\partial f_{ce}}{\partial t} + v_{||} \frac{\partial f_{ce}}{\partial x} + \frac{\partial \phi}{\partial x} \frac{\partial f_{ce}}{\partial v_{||}} = 0 ,
\end{equation}
\begin{equation}\label{m_f_se}
 \sqrt{\frac{m_{e}}{m}} \frac{\partial f_{se}}{\partial t} + v_{||} \frac{\partial f_{se}}{\partial x} + \frac{\partial \phi}{\partial x} \frac{\partial f_{se}}{\partial v_{||}} = 0 ,
\end{equation}
\begin{equation}\label{Poisson_equation_modified}
  \frac{\partial^{2} \phi}{\partial x^{2}}  = \int^{\infty}_{-\infty} f_{ce} dv_{\parallel} +\int^{\infty}_{-\infty} f_{se} dv_{||}-n .
\end{equation}
The above equations along with the equation of continuity of ions and the equation of motion for ion fluid form a system of coupled equations. The continuity equation and the momentum equation for ion fluid can be taken as
\begin{equation}\label{continuity_equation}
 \frac{\partial n}{\partial t} + \frac{\partial (nu)}{\partial x}  = 0 ,
\end{equation}
\begin{equation}\label{Equation_of_motion}
\frac{\partial u}{\partial t} + u \frac{\partial u}{\partial x} + \frac{\sigma}{n} \frac{\partial p}{\partial x} = - \frac{\partial \phi}{\partial x}.
\end{equation}
In the momentum equation (\ref{Equation_of_motion}), the pressure term has been included to get the effect of ion temperature. To make a closed system of equations, we take the following adiabatic pressure law on the basis of the assumption that the effects of viscosity, thermal conductivity and energy transfer due to collisions can be neglected.
\begin{equation}\label{Pressure_equation}
  p = n^{\gamma} .
\end{equation}
In equations (\ref{m_f_ce}) - (\ref{Pressure_equation}), $f_{ce}$, $f_{se}$, $n$, $u$, $v_{||}$, $p$, $\phi$, $x$ and $t$ are the velocity distribution function of nonthermal electrons, the velocity distribution function of isothermal electrons, the ion number density, the ion fluid velocity, the velocity of electrons in phase space, the ion pressure, the electrostatic potential, the spatial variable and time, respectively, and these quantities have been normalized by $n_{0}$ (unperturbed ion number density), $n_{0}$, $n_{0}$, $c_{s}$ $(= \sqrt{K_{B} T_{ef}/m})$, $V_{te}$ $(= \sqrt{K_{B} T_{ef}/m_{e}})$, $n_{0} K_{B} T_{i}$, $K_{B} T_{ef}/e$, $\lambda_{D}$ $(= \sqrt{K_{B} T_{ef}/4\pi n_{0} e^{2}} )$ and $\omega_{pi}^{-1} (= \sqrt{m/4\pi n_{0} e^{2}})$, where $\sigma = T_{i}/T_{ef}$ and $\gamma (=3)$ is the adiabatic index. Again, $K_{B}$ is the Boltzmann constant, $m$ is the mass of an ion, $m_{e}$ is the mass of an electron, $-e$ is the charge of an electron, $T_{i}$ is the average ion temperature and following Dalui et al. [29], $T_{ef}$ is given by the following equation:
\begin{equation}\label{equation_of_effective_temperature}
 \frac{ n_{c0} + n_{s0}}{T_{ef}} =  \frac{ n_{c0}}{T_{ce}} + \frac{ n_{s0}}{T_{se}} ,
\end{equation}
where ${n}_{c0}$, ${n}_{s0}$, $T_{ce}$ and $T_{se}$ are, respectively, unperturbed nonthermal electron number density, unperturbed isothermal electron number density, average temperature of nonthermal electrons and average temperature of isothermal electrons.

On the basis of the above mentioned normalization of the independent and dependent variables, the unperturbed velocity distribution functions of nonthermal Cairns [18] distributed electrons and isothermal electrons can be written in the following form:
\begin{equation}\label{f_c0}
  f_{c0} = \bar{n}_{c0}\sqrt{\frac{\sigma_{c}}{2\pi}}\Bigg(\frac{1+\alpha_{e}\sigma_{c}^{2}v_{||}^{4}}{1+3 \alpha_{e}} \Bigg) \exp\Big[-\frac{\sigma_{c}v_{||}^{2}}{2}\Big] ,
\end{equation}
\begin{equation}\label{f_s0}
  f_{s0} = \bar{n}_{s0} \sqrt{\frac{\sigma_{s}}{2\pi}} \exp\Big[-\frac{\sigma_{s}v_{||}^{2}}{2}\Big] ,
\end{equation}
where $\alpha_{e} (\geq 0)$ is the nonthermal parameter associated with the Cairns model [18] for hotter electron species and the expressions of $\bar{n}_{c0}$, $\bar{n}_{s0}$, $\sigma_{c}$ and $\sigma_{s}$ are given by
\begin{equation}\label{new100}
 \bar{n}_{c0}=\frac{n_{c0}}{n_{0}} ,\bar{n}_{s0}=\frac{n_{s0}}{n_{0}},\sigma_{c} = \frac{T_{ef}}{T_{ce}},\sigma_{s} = \frac{T_{ef}}{T_{se}}.
\end{equation}
Using (\ref{new100}), the equation (\ref{equation_of_effective_temperature}) and the unperturbed charge neutrality condition ($n_{c0} + n_{s0} = n_{0}$) can be written as
\begin{equation}\label{charge_neutrality_condition_1}
 \bar{n}_{c0} \sigma_{c} + \bar{n}_{s0} \sigma_{s} = 1 , \bar{n}_{c0} + \bar{n}_{s0} = 1 .
\end{equation}

Following Dalui et al. [29] and using the equation (\ref{charge_neutrality_condition_1}), we can write the expressions of $\bar{n}_{c0}$, $\bar{n}_{s0}$, $\sigma_{c}$ and $\sigma_{s}$ in the following form:
\begin{equation}\label{form_of_sigma_c1}
  \bar{n}_{s0} = \frac{n_{sc}}{1+n_{sc}},\bar{n}_{c0}=\frac{1}{1+n_{sc}},
\end{equation}
\begin{equation}\label{form_of_sigma_c2}
  \sigma_{s} = \frac{1+n_{sc}}{\sigma_{sc}+n_{sc}},\sigma_{c} = \sigma_{sc}\frac{1+n_{sc}}{\sigma_{sc}+n_{sc}} , 
\end{equation}
where 
\begin{equation}\label{n_sc_and_sigma_sc}
  n_{sc} = \frac{n_{s0}}{n_{c0}} \mbox{        and         } \sigma_{sc} = \frac{T_{se}}{T_{ce}}.
\end{equation}
Therefore, $n_{sc}$, $\sigma_{sc}$, $\sigma$ and $\alpha_{e}$ are the four basic parameters of the present problem.

In the present paper, we follow the method of Ott and Sudan [45] to study the effect of electron Landau damping on the IA solitary wave in an unmagnetized plasma including weak nonlinearity and weak dispersion. So, we replace $\displaystyle \sqrt{m_{e}/m}$ by $\displaystyle \epsilon \sqrt{m_{e}/m}$ to indicate the smallness of the quantity $\displaystyle \sqrt{m_{e}/m}$, where $\epsilon$ is a small parameter. So, writing $\displaystyle \alpha_{1} = \sqrt{m_{e}/m}$ and replacing $\displaystyle \sqrt{m_{e}/m}$ by $\displaystyle \epsilon \alpha_{1}$, the equations (\ref{m_f_ce}) and (\ref{m_f_se}) can be written in the following form:
\begin{equation}\label{f_ce}
 \alpha_{1}\epsilon \frac{\partial f_{ce}}{\partial t} + v_{||} \frac{\partial f_{ce}}{\partial x} + \frac{\partial \phi}{\partial x} \frac{\partial f_{ce}}{\partial v_{||}} = 0 ,
\end{equation}
\begin{equation}\label{f_se}
 \alpha_{1}\epsilon \frac{\partial f_{se}}{\partial t} + v_{||} \frac{\partial f_{se}}{\partial x} + \frac{\partial \phi}{\partial x} \frac{\partial f_{se}}{\partial v_{||}} = 0 .
\end{equation}
Again, using (\ref{Pressure_equation}), the momentum equation (\ref{Equation_of_motion}) can be written as
\begin{equation}\label{Equation_of_motion_modified}
\frac{\partial u}{\partial t} + u \frac{\partial u}{\partial x} +\gamma \sigma n^{\gamma-2} \frac{\partial n}{\partial x} = - \frac{\partial \phi}{\partial x}.
\end{equation}
Therefore, the equations (\ref{f_ce}), (\ref{f_se}), (\ref{Poisson_equation_modified}), (\ref{continuity_equation}) and (\ref{Equation_of_motion_modified}) are the basic equations to study the nonlinear behaviour of IA waves including the effect of electron Landau damping.

\begin{center}
\textbf{3. DERIVATION OF DIFFERENT EVOLUTION EQUATIONS}
\end{center}

To derive different nonlinear evolution equations including the kinetic  effect of electrons on IA waves propagating along $x$-axis, we consider the following stretching of the space coordinate and time:
\begin{equation}\label{Stretching}
  \xi = \epsilon^{\frac{1}{2}} (x-Vt) , \tau = \epsilon^{\frac{3}{2}} t ,
\end{equation}
where $V$ is a constant and $\epsilon$ is a small parameter.

\begin{center}
\textbf{3.1. KdV equation including the effect of Landau damping}
\end{center}

To derive the KdV equation including the effect of Landau damping, we take the following perturbation expansions of the dependent variables:
\begin{equation}\label{KdV_Perturbation_of_s}
  \Lambda = \Lambda^{(0)} + \sum_{i=1}^{\infty} \epsilon^{i} \Lambda^{(i)}(\xi,\tau) ,
\end{equation}
where $\Lambda$ = $n$, $u$, $\phi$, $f_{ce}$ and $f_{se}$ with ($n^{(0)}$,$u^{(0)}$, $\phi^{(0)}$, $f_{ce}^{(0)}$, $f_{se}^{(0)}$) = ($1$, $0$, $0$, $f_{c0}$, $f_{s0}$).

Substituting (\ref{Stretching}) and  (\ref{KdV_Perturbation_of_s}) 
into the equations (\ref{f_ce}), (\ref{f_se}), (\ref{Poisson_equation_modified}), (\ref{continuity_equation}) and (\ref{Equation_of_motion_modified}) and collecting the terms of different powers of $\epsilon $ on both sides of each equation, we get a sequence of equations and from this sequence of equations, we get the following nonlinear evolution equation:
\begin{eqnarray}\label{KdV_equation}
   \frac{\partial \phi^{(1)}}{\partial \tau} + AB_{1} \phi^{(1)} \frac{\partial \phi^{(1)}}{\partial \xi} + \frac{1}{2}A \frac{\partial^{3} \phi^{(1)}}{\partial \xi^3}+ \frac{1}{2}AE \alpha_{1} \mathcal{P} \int^{\infty}_{-\infty} \frac{\partial \phi^{(1)}}{\partial \xi'} \frac{d \xi'}{\xi-\xi'} = 0 ,
\end{eqnarray}
where we have used the same procedure of Bandyopadhyay and Das [48] to derive the equation (\ref{KdV_equation}). 

The coefficients $A$, $B_{1}$, $E$ are given by
\begin{equation}\label{A}
	A = \frac{1}{V}(V^{2}-\sigma \gamma)^{2}  , 
\end{equation}
\begin{equation}\label{B}
	B_{1} = \frac{1}{2} \bigg[ \frac{3V^{2}+\sigma \gamma (\gamma-2)}{(V^{2}-\sigma \gamma)^{3}} - \big(\bar{n}_{c0} \sigma_{c}^2 + \bar{n}_{s0} \sigma_{s}^2\big)  \bigg],
\end{equation}
\begin{equation}\label{E}
	E = \frac{V}{\sqrt{2\pi}} \Big[	\bar{n}_{c0} \sigma_{c}^{3/2} \big(1-\frac{3}{4}\beta_{e}\big) + \bar{n}_{s0} \sigma_{s}^{3/2}  \Big]. 
\end{equation}
The constant $V$ is given by
\begin{equation}\label{dispersion_relation}
   (V^{2}-\sigma \gamma)(1 - \bar{n}_{c0} \sigma_{c}\beta_{e} )=1  ,
\end{equation}
where $\displaystyle \beta_{e}=\frac{4 \alpha_{e}}{1+3\alpha_{e}}$ and the physically admissible range of $\beta_{e}$ is $0 \leq \beta_{e} \leq \frac{4}{7}$ although mathematically $\beta_{e}$ is restricted by the inequality: $0 \leq \beta_{e} \leq \frac{4}{3}$. So, instead of $\alpha_{e}$ we can take $\beta_{e}$ as nonthermal parameter and consequently the new set of basic parameters of the present plasma system are $n_{sc}$, $\sigma_{sc}$, $\sigma$ and $\beta_{e}$. 

If we neglect electron to ion mass ratio, i.e., if we set $\alpha_{1}=0$, then the nonlinear evolution equation (\ref{KdV_equation}) simply reduces to the well known KdV equation.
 
Equation (\ref{KdV_equation}) describes the propagation of weakly nonlinear and weakly dispersive IA solitary waves in a multi-species collisionless unmagnetized plasma consisting of nonthermal and isothermal electrons including the effect of Landau damping.

From equation (\ref{KdV_equation}), we see that the nonlinearity of the IA wave is only due to the second term of (\ref{KdV_equation}), i.e., $AB_{1}$ is responsible for the nonlinearity of the system. When $AB_{1}=0$, i.e., $B_{1}=0$ (as $A \neq 0$ for any set of physically admissible values of the parameters of the system), it is not possible to discuss  the nonlinear behavior of IA waves with the help of the evolution equation (\ref{KdV_equation}).

\begin{figure}
\begin{center}
\includegraphics{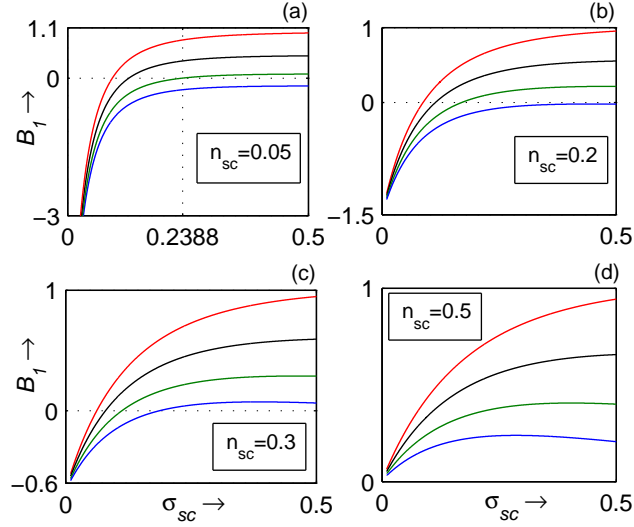}
  \caption{\label{B1_Vs_sigma_sc_diff_beta_e_subplot} $B_{1}$ is plotted against $\sigma_{sc}$ for $\gamma=3$, $\sigma=0.001$ and for (a) $n_{sc}=0.05$, (b) $n_{sc}=0.2$, (c) $n_{sc}=0.3$ and (d) $n_{sc}=0.5$. Red, black, green and blue curves of each figure correspond to $\beta_{e}=0$, $\beta_{e}=0.2$, $\beta_{e}=0.4$ and $\beta_{e}=0.57$ respectively. Figures \ref{B1_Vs_sigma_sc_diff_beta_e_subplot}(a), \ref{B1_Vs_sigma_sc_diff_beta_e_subplot}(b) and \ref{B1_Vs_sigma_sc_diff_beta_e_subplot}(c) show the existence of points  $\sigma_{sc}^{(c)}$ such that $B_{1}=0$ for some values of $\beta_{e}$ whereas figure \ref{B1_Vs_sigma_sc_diff_beta_e_subplot}(d) shows that $B_{1}>0$ for all values of $\beta_{e}$ and for all $\sigma_{sc}$ lying within the interval $(0,0.5)$. In particular, for $n_{sc}=0.05$, $\beta_{e}=0.4$, the value of $\sigma_{sc}^{(c)}$ is $0.2388$ (approx.).}
\end{center}
\end{figure}

In figure \ref{B1_Vs_sigma_sc_diff_beta_e_subplot}, $B_{1}$ is plotted against $\sigma_{sc}$ for $\gamma=3$, $\sigma=0.001$ and for (a) $n_{sc}=0.05$, (b) $n_{sc}=0.2$, (c) $n_{sc}=0.3$ and (d) $n_{sc}=0.5$. Here, red, black, green and blue curves of each figure correspond to $\beta_{e}=0$, $\beta_{e}=0.2$, $\beta_{e}=0.4$ and $\beta_{e}=0.57$ respectively. From figures \ref{B1_Vs_sigma_sc_diff_beta_e_subplot}(a), \ref{B1_Vs_sigma_sc_diff_beta_e_subplot}(b) and \ref{B1_Vs_sigma_sc_diff_beta_e_subplot}(c), we see that there exists a value $\sigma_{sc}^{(c)}$ of $\sigma_{sc}$ such that $B_{1}=0$ at $\sigma_{sc}$=$\sigma_{sc}^{(c)}$, and more specifically, $B_{1} < 0 $ for $\sigma_{sc} < \sigma_{sc}^{(c)}$ and $B_{1} > 0 $ for $\sigma_{sc} > \sigma_{sc}^{(c)}$. Again, from figure \ref{B1_Vs_sigma_sc_diff_beta_e_subplot}(d), we see that $B_{1}>0$ for all values of $\beta_{e}$. From figure \ref{B1_Vs_sigma_sc_diff_beta_e_subplot}, it is evident that there exists a region $R_{I}=\{( n_{sc}, \sigma_{sc}, \beta_{e}):B_{1}( n_{sc}, \sigma_{sc}, \beta_{e}) \neq 0\}$ such that each point of $R_{I}$ satisfies the condition $B_{1} ( n_{sc}, \sigma_{sc}, \beta_{e}) \neq 0$. On the other hand there must exist a collection of points from the entire parameter space such that every point of the collection must satisfy the equation $B_{1} ( n_{sc}, \sigma_{sc}, \beta_{e}) = 0$ and consequently for these values of the parameters $n_{sc}$, $\sigma_{sc}$ and $\beta_{e}$ we cannot use the KdV like evolution equation to investigate the effect of Landau damping on IA solitary waves. To confirm the existence of a region $R_{II}=\{( n_{sc}, \sigma_{sc}, \beta_{e}):B_{1}( n_{sc}, \sigma_{sc}, \beta_{e})=0\}$ in the entire parameter space, we consider the following figures in different parameter planes.     

\begin{figure}
\begin{center}
\includegraphics{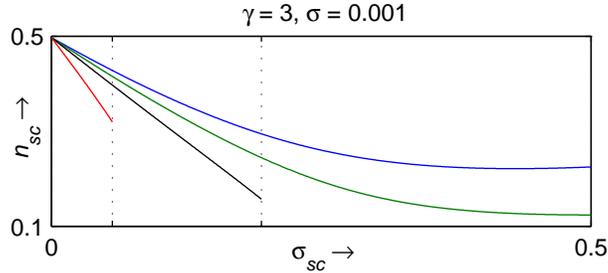}
  \caption{\label{Final_B1_eq_zero} $n_{sc}$ is plotted against $\sigma_{sc}$ when $B_{1}=0$ for different values of $\beta_{e}$. For every value of $\beta_{e}$, we have a curve in the $\sigma_{sc}$ - $n_{sc}$ parameter plane and at every point on this curve we get a value of $\sigma_{sc}$ as well as a value of $n_{sc}$, and finally for these values of $\beta_{e}$, $\sigma_{sc}$ and $n_{sc}$, the equation $B_{1}=0$ holds good. Red, black, green and blue curves correspond to $\beta_{e}=0$, $\beta_{e}=0.4$, $\beta_{e}=0.5$ and $\beta_{e}=0.57$ respectively.}
\end{center}
\end{figure}

Now, it is simple to check that $B_{1}$ is a function of $n_{sc}$, $\sigma_{sc}$ and $\beta_{e}$ for any prescribed value of $\sigma$ and $\gamma$, i.e., $B_{1}=B_{1} ( n_{sc}, \sigma_{sc}, \beta_{e})$. Throughout this paper we take $\gamma=3$ and $\sigma = 0.001$, then all the coefficients $A$, $B_{1}$, $E$ can be regarded as functions of $n_{sc}$, $\sigma_{sc}$ and $\beta_{e}$. Therefore, $B_{1}$ is a function of $\sigma_{sc}$ and $n_{sc}$ for any given value of $\beta_{e}$, and consequently, $B_{1}=0$ gives a functional relationship between $\sigma_{sc}$ and $n_{sc}$. This functional relationship between $\sigma_{sc}$ and $n_{sc}$ is plotted in figure \ref{Final_B1_eq_zero} when $B_{1} ( n_{sc}, \sigma_{sc}, \beta_{e})=0$ for different values of $\beta_{e}$. Here, red, black, green and blue curves correspond to $\beta_{e}=0$, $\beta_{e}=0.4$, $\beta_{e}=0.5$ and $\beta_{e}=0.57$ respectively. From this figure, we see that the interval of existence of $\sigma_{sc}$ increases with increasing $\beta_{e}$ when $B_{1} ( n_{sc}, \sigma_{sc}, \beta_{e})=0$.

\begin{figure}
\begin{center}
\includegraphics{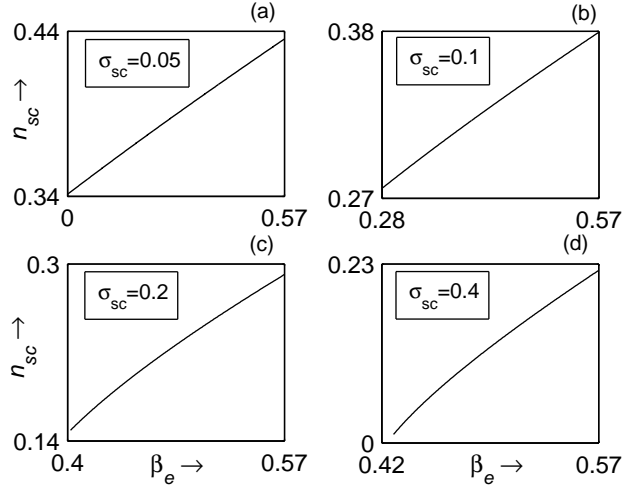}
  \caption{\label{B1_eq_zero_1_n_sc_vs_beta_e_subplot} $n_{sc}$ is plotted against $\beta_{e}$ when $B_{1}=0$ for $\sigma=0.001$ and for different values of $\sigma_{sc}$. For every value of $\sigma_{sc}$, we have a curve in the $\beta_{e}$ - $n_{sc}$ parameter plane and at every point on this curve we get a value of $\beta_{e}$ as well as a value of $n_{sc}$, and finally for these values of $\sigma_{sc}$, $\beta_{e}$ and $n_{sc}$, the equation $B_{1}=0$ holds good.}
\end{center}
\end{figure}

Again, from the equation $B_{1} ( n_{sc}, \sigma_{sc}, \beta_{e})=0$, we get a functional relationship between $n_{sc}$ and $\beta_{e}$ for any given value of $\sigma_{sc}$. In figure \ref{B1_eq_zero_1_n_sc_vs_beta_e_subplot}, $n_{sc}$ is plotted against $\beta_{e}$ when $B_{1} ( n_{sc}, \sigma_{sc}, \beta_{e})=0$ for (a) $\sigma_{sc}=0.05$, (b) $\sigma_{sc}=0.1$, (c) $\sigma_{sc}=0.2$ and (d) $\sigma_{sc}=0.4$. From this figure, we see that the interval of existence of $\beta_{e}$ decreases with increasing  $\sigma_{sc}$ when $B_{1} ( n_{sc}, \sigma_{sc}, \beta_{e})=0$. 

\begin{figure}
\begin{center}
\includegraphics{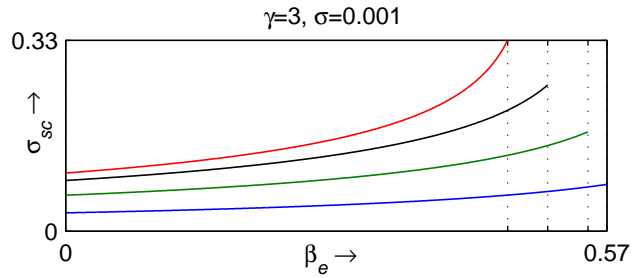}
  \caption{\label{B1_eq_zero_1_sigma_sc_vs_beta_e} $\sigma_{sc}$ is plotted against $\beta_{e}$ when $B_{1}=0$ for different values of $n_{sc}$. For a fixed value of $n_{sc}$, we have a curve in the $\beta_{e}$ - $\sigma_{sc}$ parameter plane and at every point on this curve we get a value of $\beta_{e}$ as well as a value of $\sigma_{sc}$, and finally for these values of $n_{sc}$, $\beta_{e}$ and $\sigma_{sc}$, the equation $B_{1}=0$ holds good. Red, black, green and blue curves correspond to $n_{sc}=0.1$, $n_{sc}=0.2$, $n_{sc}=0.3$ and $n_{sc}=0.4$ respectively.}
\end{center}
\end{figure}

Similarly, when $B_{1} ( n_{sc}, \sigma_{sc}, \beta_{e})=0$, we get a functional relationship between $\sigma_{sc}$ and $\beta_{e}$ for any fixed value of $n_{sc}$. When $B_{1} ( n_{sc}, \sigma_{sc}, \beta_{e})=0$, then the functional relation between $\sigma_{sc}$ and $\beta_{e}$ is plotted in figure \ref{B1_eq_zero_1_sigma_sc_vs_beta_e} for different values of $n_{sc}$ with $\sigma=0.001$. Red, black, green and blue curves correspond to $n_{sc}=0.1$, $n_{sc}=0.2$, $n_{sc}=0.3$ and $n_{sc}=0.4$ respectively. From this figure, we see that the interval of existence of $\beta_{e}$ increases with increasing  $n_{sc}$ whereas $\sigma_{sc}$ decreases with increasing $n_{sc}$ for any fixed $\beta_{e}$. 

So, figures \ref{B1_Vs_sigma_sc_diff_beta_e_subplot} - \ref{B1_eq_zero_1_sigma_sc_vs_beta_e} confirm the existence of a region $R_{II}=\{( n_{sc}, \sigma_{sc}, \beta_{e}):B_{1}( n_{sc}, \sigma_{sc}, \beta_{e})=0\}$ in the parameter space such that each point of $R_{II}$ satisfies the equation $B_{1} ( n_{sc}, \sigma_{sc}, \beta_{e})=0$. Therefore, for $B_{1} ( n_{sc}, \sigma_{sc}, \beta_{e})=0$ or for $( n_{sc}, \sigma_{sc}, \beta_{e}) \in R_{II}$, it is necessary to modify the KdV - like evolution equation to investigate the effect of Landau damping on IA solitary waves.

\begin{center}
\textbf{3.2. MKdV equation including the Landau damping effect}
\end{center}

When $B_{1}=0$, we take the following perturbation expansions of the dependent variables:
\begin{equation}\label{mKdV_Perturbation_of_s}
  \Lambda = \Lambda^{(0)} + \sum_{i=1}^{\infty} \epsilon^{\frac{i}{2}} \Lambda^{(i)}(\xi,\tau) ,
\end{equation}
where $\Lambda$ = $n$, $u$, $\phi$, $f_{ce}$ and $f_{se}$ with ($n^{(0)}$,$u^{(0)}$, $\phi^{(0)}$, $f_{ce}^{(0)}$, $f_{se}^{(0)}$) = ($1$, $0$, $0$, $f_{c0}$, $f_{s0}$).

Substituting (\ref{Stretching}) and (\ref{mKdV_Perturbation_of_s}) 
into the equations (\ref{f_ce}), (\ref{f_se}), (\ref{Poisson_equation_modified}), (\ref{continuity_equation}) and (\ref{Equation_of_motion_modified}) and collecting the terms of different powers of $\epsilon $, we get a sequence of equations and from this sequence of equations, following Bandyopadhyay and Das [48], we get the following nonlinear evolution equation:
\begin{eqnarray}\label{mKdV_KdV_equation}
   \frac{\partial \phi^{(1)}}{\partial \tau} + AB_{2} [\phi^{(1)}]^{2} \frac{\partial \phi^{(1)}}{\partial \xi} + \frac{1}{2}A \frac{\partial^{3} \phi^{(1)}}{\partial \xi^3}+ \frac{1}{2}AE \alpha_{1}\mathcal{P} \int^{\infty}_{-\infty} \frac{\partial \phi^{(1)}}{\partial \xi'} \frac{d \xi'}{\xi-\xi'} = 0.
\end{eqnarray}
Here, it is important to mention that the condition $B_{1}=0$ has been used to eliminate the term $AB_{1} \frac{\partial (\phi^{(1)}\phi^{(2)})}{\partial \xi}$ from the final form of (\ref{mKdV_KdV_equation}). The expressions of $A$, $E$, $V$ are given by (\ref{A}), (\ref{E}), (\ref{dispersion_relation}) respectively, and the expression of  $B_{2}$ can be written as  
\begin{equation}\label{mKdV_B_2}
		B_{2} = \frac{1}{4} \Big[H_{2}  - \big(\bar{n}_{c0} \sigma_{c}^3 (1+3\beta_{e}) + \bar{n}_{s0} \sigma_{s}^3 \big) \Big] ,
\end{equation}
where 
\begin{eqnarray}\label{mKdV_M_2}
		 H_{2} = \frac{1}{(V^{2}-\sigma \gamma)^{5}} \Big[15V^{4} + \sigma \gamma (\gamma^{2}+13\gamma-18)V^{2}+ \sigma^{2} \gamma^{2} (\gamma-2)(2\gamma-3)\Big].
\end{eqnarray}
If $\alpha_{1}=0$, then the nonlinear evolution equation (\ref{mKdV_KdV_equation}) simply reduces to the well known MKdV equation.

From equation (\ref{mKdV_KdV_equation}), we see that the nonlinearity of the IA wave is only due to the second term of (\ref{mKdV_KdV_equation}). So, the equation (\ref{mKdV_KdV_equation}) describes the nonlinear dynamics of IA waves when $B_{1}=0$ and $B_{2} \neq 0$. 

\begin{figure}
\begin{center}
\includegraphics{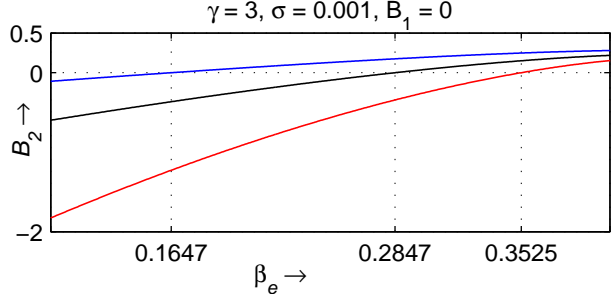}
  \caption{\label{Final_sample_B2_vs_beta_e_when_B1_eq_zero_diff_n_sc} $B_{2}$ is plotted against $\beta_{e}$ when $B_{1}=0$ for different values of $n_{sc}$, i.e., the solution of the equation $B_{1}=0$ for the unknown $\sigma_{sc}$ gives $\sigma_{sc}$ as a function of $\beta_{e}$ and consequently one can express $B_{2}$ as a function of $\beta_{e}$, this $B_{2}$ is plotted against $\beta_{e}$.  Red, black and blue curves correspond to $n_{sc}=0.02$, $n_{sc}=0.05$ and $n_{sc}=0.08$ respectively. This figure shows the existence of a point $\beta_{e}^{(c)}$ of $\beta_{e}$ where $B_{2}=0$. In particular, for $n_{sc}=0.05$, the value of $\beta_{e}^{(c)}$ is approximately 0.2847.}
\end{center}
\end{figure}

Now, in figure \ref{Final_sample_B2_vs_beta_e_when_B1_eq_zero_diff_n_sc}, $B_{2}$ is plotted against $\beta_{e}$ when $B_{1}=0$ for $\gamma=3$ and $\sigma=0.001$, and for different values of $n_{sc}$. In fact, for given values of 
$\gamma$, $\sigma$ and $n_{sc}$, $B_{1}$ is a function of $\sigma_{sc}$ and $\beta_{e}$ only and consequently if we solve the equation $B_{1}=0$ with respect to the unknown $\sigma_{sc}$, we get $\sigma_{sc}$ as a function of $\beta_{e}$. If we put all the values of $\gamma$, $\sigma$, $n_{sc}$ and $\sigma_{sc}$ in the expression of $B_{2}$, we get $B_{2}$ as a function of $\beta_{e}$. This $B_{2}$ is plotted against $\beta_{e}$ in figure \ref{Final_sample_B2_vs_beta_e_when_B1_eq_zero_diff_n_sc}.  Here red, black and blue curves correspond to $n_{sc}=0.02$, $n_{sc}=0.05$ and $n_{sc}=0.08$ respectively. This figure clearly shows that there exists a value $\beta_{e}^{(c)}$ of $\beta_{e}$ such that $B_{2}=0$ at $\beta_{e}=\beta_{e}^{(c)}$ and more specifically, $B_{2} < 0 $ for $\beta_{e} < \beta_{e}^{(c)}$, $B_{2} > 0 $ for $\beta_{e} > \beta_{e}^{(c)}$ and $B_{2}=0$ at $\beta_{e}=\beta_{e}^{(c)}$. In particular, for $n_{sc}=0.05$, the value of $\beta_{e}^{(c)}$ is approximately equal to $0.2847$. Therefore, there exist points $(n_{sc}, \sigma_{sc}, \beta_{e})$ in the parameter space such that $B_{1}=B_{2}=0$. So, now it is necessary to divide the region $R_{II}$ into two regions $R_{II}^{(a)}$ and $R_{II}^{(b)}$ such that $R_{II}^{(a)}=\{(n_{sc}, \sigma_{sc}, \beta_{e}):B_{1}=0 \mbox{     and     } B_{2} \neq 0\}$ and $R_{II}^{(b)}=\{(n_{sc}, \sigma_{sc}, \beta_{e}):B_{1}= B_{2} = 0\}$.

\begin{figure}
\begin{center}
\includegraphics{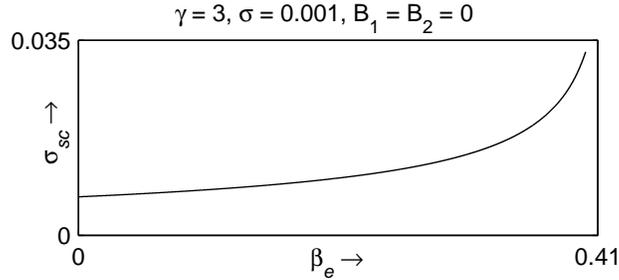}
  \caption{\label{sample_sigma_sc_vs_beta_e_when_B1_B2_eq_zero} $\sigma_{sc}$ is plotted against $\beta_{e}$ when $B_{1}=B_{2}=0$. Here $B_{1}$ and $B_{2}$ both are the functions of $n_{sc}$, $\sigma_{sc}$ and $\beta_{e}$. Therefore, solving the equation $B_{1}=0$ for the unknown $n_{sc}$, we get $n_{sc}$ as a function of $\sigma_{sc}$ and $\beta_{e}$. If we put this solution for $n_{sc}$ in the expression of $B_{2}$, we get $B_{2}$ as a function of $\sigma_{sc}$ and $\beta_{e}$. Finally, the equation $B_{2}=0$ gives $\sigma_{sc}$ as a function of $\beta_{e}$. We plot this solution $\sigma_{sc}$ against $\beta_{e}$ and along this curve, we have $B_{1}=B_{2}=0$. }
\end{center}
\end{figure}

We see that the equation (\ref{mKdV_KdV_equation}) is free from any nonlinear effect when $B_{1}=B_{2}=0$, i.e., if $(n_{sc}, \sigma_{sc}, \beta_{e}) \in R_{II}^{(b)}$. To explain the existence of the region $R_{II}^{(b)}$, we consider figure \ref{sample_sigma_sc_vs_beta_e_when_B1_B2_eq_zero}. Now, it is simple to check that $B_{1}$ and $B_{2}$ are the functions of $n_{sc}$, $\sigma_{sc}$ and $\beta_{e}$ for any prescribed value of $\sigma$ and $\gamma$, i.e., $B_{1}=B_{1} ( n_{sc}, \sigma_{sc}, \beta_{e})$ and $B_{2}=B_{2} ( n_{sc}, \sigma_{sc}, \beta_{e})$ for any given value of $\sigma$ and $\gamma$. We have mentioned earlier that throughout this paper we take $\gamma=3$ and $\sigma = 0.001$. Now, for given values of $\beta_{e}$ and $\sigma_{sc}$, $B_{1} (n_{sc}, \sigma_{sc}, \beta_{e})=0$ gives an equation for the unknown $n_{sc}$ and consequently,  $B_{1}=0$ gives a real solution for $n_{sc}$. Let $n_{sc}=n_{sc}(\beta_{e}, \sigma_{sc})$ be the physically admissible real solution of the equation $B_{1} (n_{sc}, \sigma_{sc}, \beta_{e})=0$, i.e., the physically admissible real solution $n_{sc}$ of the equation $B_{1} (n_{sc}, \sigma_{sc}, \beta_{e})=0$ can be considered as a function of $\beta_{e}$ and $\sigma_{sc}$. If we put this value of $n_{sc}(\beta_{e}, \sigma_{sc})$ in the expression of $B_{2} (n_{sc}, \sigma_{sc}, \beta_{e})$ then the function $B_{2}$ is reduced to a function of $\beta_{e}$ and $\sigma_{sc}$ only, i.e., $B_{2}=B_{2} (\beta_{e}, \sigma_{sc})$. Again, $B_{2}=B_{2} (\beta_{e}, \sigma_{sc})=0$ gives a functional relationship between $\sigma_{sc}$ and $\beta_{e}$. This functional relationship between $\sigma_{sc}$ and $\beta_{e}$ is plotted in figure \ref{sample_sigma_sc_vs_beta_e_when_B1_B2_eq_zero}, for fixed values of the other parameters, i.e., in figure \ref{sample_sigma_sc_vs_beta_e_when_B1_B2_eq_zero}, $\sigma_{sc}$ is plotted against the nonthermal parameter $\beta_{e}$ when $B_{1}=0$ and $B_{2}=0$. Figure \ref{sample_sigma_sc_vs_beta_e_when_B1_B2_eq_zero} shows a variation of $\sigma_{sc}$ against the nonthermal parameter $\beta_{e}$ in $\beta_{e}$-$\sigma_{sc}$ parameter plane when $B_{1}=B_{2}=0$. This figure shows the existence of a curve in the $\beta_{e}$-$\sigma_{sc}$ parameter plane along which $B_{1}=0$ and $B_{2}=0$. Different values of $\sigma$ will give different curves in the $\beta_{e}$-$\sigma_{sc}$ parameter plane along which $B_{1}=0$ and $B_{2}=0$. Therefore, the existence of region $R_{II}^{(b)}$ in the parameter space is confirmed and consequently, in this region of parameter space, it is not possible to describe the nonlinear dynamics of IA waves either by the KdV like equation (\ref{KdV_equation}) or by the MKdV like equation (\ref{mKdV_KdV_equation}). Therefore, for $B_{1}=B_{2}=0$, a further modification of the evolution equation (\ref{mKdV_KdV_equation}) is necessary to study the effect of Landau damping on IA solitary waves. In the next subsection, we have derived a new evolution equation when the conditions $B_{1}=0$ and $B_{2}=0$ hold simultaneously. 

\begin{center}
\textbf{3.3. FMKdV equation including the Landau damping effect}
\end{center} 

To derive the FMKdV equation including the effect of Landau damping when the conditions $B_{1}=0$ and $B_{2}=0$ hold simultaneously, we take the following perturbation expansions of the dependent variables:
\begin{equation}\label{fmKdV_Perturbation_of_s}
  \Lambda = \Lambda^{(0)} + \sum_{i=1}^{\infty} \epsilon^{\frac{i}{3}} \Lambda^{(i)}(\xi,\tau) ,
\end{equation}
where $\Lambda$ = $n$, $u$, $\phi$, $f_{ce}$ and $f_{se}$ with ($n^{(0)}$,$u^{(0)}$, $\phi^{(0)}$, $f_{ce}^{(0)}$, $f_{se}^{(0)}$) = ($1$, $0$, $0$, $f_{c0}$, $f_{s0}$).

Substituting (\ref{Stretching}) and (\ref{fmKdV_Perturbation_of_s}) into the equations (\ref{f_ce}), (\ref{f_se}), (\ref{Poisson_equation_modified}), (\ref{continuity_equation}) and (\ref{Equation_of_motion_modified}) and collecting the terms of different powers of $\epsilon $ on both sides of each equation, we get a sequence of equations.

\begin{center}
\textbf{3.3.1. Equations for ion fluid at the order $\epsilon^{5/6}$}
\end{center}

At the order $\epsilon^{5/6}$, solving the equation of continuity and the equation of motion of ion fluid for the unknowns $n^{(1)}$ and $u^{(1)}$, we get 
\begin{equation}\label{fmKdV_sol_n1_and_u1}
  n^{(1)} = \frac{1}{V^{2}-\sigma \gamma} \phi^{(1)}, u^{(1)} = \frac{V}{V^{2}-\sigma \gamma} \phi^{(1)} .
\end{equation}

\begin{center}
\textbf{3.3.2. Vlasov-Boltzmann equation at the order $\epsilon^{5/6}$}
\end{center} 

The Vlasov-Boltzmann equation of nonthermal electrons at the order $\epsilon^{5/6}$ is
\begin{equation}\label{fmKdV_Vlasov_f_ce_11}
  v_{||}\frac{\partial f_{ce}^{(1)}}{\partial \xi} + \frac{\partial \phi^{(1)}}{\partial \xi} \frac{\partial f_{c0}}{\partial v_{||}} = 0 .
\end{equation}
The above equation does not have a unique solution and consequently to get the unique solution of the equation (\ref{fmKdV_Vlasov_f_ce_11}), we follow the method of Ott \& Sudan [45]. This method suggests to add an extra higher order time derivative term $\epsilon^{17/6}\alpha_{1}\frac{\partial f_{ce}^{(1)}}{\partial \tau}$ with the Vlasov-Boltzmann equation at the order $\epsilon^{5/6}$. So, the equation (\ref{fmKdV_Vlasov_f_ce_11}) can be written in the following form:
\begin{equation}\label{fmKdV_Vlasov_f_ce_115}
    \alpha_{1} \epsilon^{2}\frac{\partial f_{ce\epsilon}^{(1)}}{\partial \tau} + v_{||}\frac{\partial f_{ce\epsilon}^{(1)}}{\partial \xi} + \frac{\partial \phi^{(1)}}{\partial \xi} \frac{\partial f_{c0}}{\partial v_{||}}  = 0 ,
\end{equation}
where $f_{ce}^{(1)}$ is replaced by $f_{ce\epsilon}^{(1)}$ and one can get $f_{ce}^{(1)}$ from the solution of the above equation by considering the following relation for $j=1$.
\begin{equation}\label{f_ce_epsilon_j}
   f_{ce}^{(j)}=\lim_{\epsilon \to 0} f_{ce\epsilon}^{(j)} ~,~j=1,2,3,\cdots .
\end{equation}
 To solve (\ref{fmKdV_Vlasov_f_ce_115}), we have assumed that the time dependence of any perturbed quantity is of the form $\exp(i \omega \tau)$ and we can write the equation (\ref{fmKdV_Vlasov_f_ce_115}) as
 \begin{equation}\label{fmKdV_Vlasov_f_ce_tau_dependence_1}
  i \alpha_{1} \omega \epsilon^{2} f_{ce\epsilon}^{(1)}  + v_{||}\frac{\partial f_{ce\epsilon}^{(1)}}{\partial \xi} + \frac{\partial \phi^{(1)}}{\partial \xi} \frac{\partial f_{c0}}{\partial v_{||}} 
 = 0.
\end{equation}
Now, taking the Fourier transform of this equation with respect to $\xi$, we get
\begin{equation}\label{Vlasov_f_ce41}
 \widetilde{f}_{ce\epsilon}^{(1)} = -2\frac{\partial f_{c0}}{\partial v_{||}^{2}}\frac{sv_{||}}{sv_{||}+\alpha_{1} \omega \epsilon^{2}} \widetilde{\phi}^{(1)} ,
\end{equation} 
where the Fourier transform of $g$ with respect to $\xi$ is defined as 
\begin{equation}\label{Fourier_transform_1}
 \widetilde{g}=\frac{1}{\sqrt{2\pi}} \int^{\infty}_{-\infty} g(\xi) e^{-is\xi} d\xi .
\end{equation}
Again, using the Landau prescription to resolve the singularities involved, the equation (\ref{Vlasov_f_ce41}) can be written as
\begin{eqnarray}\label{Vlasov_f_ce5}
 \widetilde{f}_{ce\epsilon}^{(1)} = -2\frac{\partial f_{c0}}{\partial v_{||}^{2}}\Big\{sv_{||}\mathcal{P} \frac{1}{sv_{||}+\alpha_{1} \omega \epsilon^{2}} +  i \pi sv_{||} \delta(sv_{||}+\alpha_{1} \omega \epsilon^{2})\Big\} \widetilde{\phi}^{(1)}.
\end{eqnarray}
Taking limit $\epsilon \rightarrow 0$, we get
\begin{equation}\label{Vlasov_f_ce5_modified}
 \widetilde{f}_{ce}^{(1)} = -2\frac{\partial f_{c0}}{\partial v_{||}^{2}}\Big\{sv_{||}\mathcal{P} \frac{1}{sv_{||}}  
 +i \pi sv_{||} \delta(sv_{||})\Big\} \widetilde{\phi}^{(1)},
\end{equation}
where we have used the relation (\ref{f_ce_epsilon_j}) for $j=1$.

Now, using the relations $x\mathcal{P}(1/x)=1$ and $x\delta(x)=0$, equation (\ref{Vlasov_f_ce5_modified}) can be simplified as
\begin{equation}\label{Vlasov_f_ce5_modified_modified}
 \widetilde{f}_{ce}^{(1)} = -2\frac{\partial f_{c0}}{\partial v_{||}^{2}} \widetilde{\phi}^{(1)}.
\end{equation}
Taking Fourier inversion of the above equation, we get
\begin{equation}\label{Vlasov_f_ce6}
 f_{ce}^{(1)} = - 2\frac{\partial f_{c0}}{\partial v_{||}^{2}} \phi^{(1)}.
\end{equation}

Similarly, considering the Vlasov-Boltzmann equation of isothermal electrons at the order $\epsilon^{5/6}$,  we get
\begin{equation}\label{Vlasov_f_se6}
 f_{se}^{(1)} = - 2\frac{\partial f_{s0}}{\partial v_{||}^{2}} \phi^{(1)} .
\end{equation}

\begin{center}
\textbf{3.3.3. Poisson equation at the order $\epsilon^{\frac{1}{3}}$}
\end{center}

From the Poisson equation at the order $\epsilon^{\frac{1}{3}}$, we get
\begin{equation}\label{Poisson_equation_modified_oreder1}
  n^{(1)}  = \int^{\infty}_{-\infty} f_{ce}^{(1)} dv_{\parallel} +\int^{\infty}_{-\infty} f_{se}^{(1)} dv_{||}.
\end{equation}
Using (\ref{Vlasov_f_ce6}) and (\ref{Vlasov_f_se6}), the above equation can be written in the following form:
\begin{equation}\label{sol_nce1_and_nse3}
  n^{(1)} = (1 - \bar{n}_{c0} \sigma_{c}\beta_{e} ) \phi^{(1)} .
\end{equation}
Using this equation and the first equation of (\ref{fmKdV_sol_n1_and_u1}), we get the equation (\ref{dispersion_relation}). Therefore, the Poisson equation at the order $\epsilon^{\frac{1}{3}}$ gives the dispersion relation (\ref{dispersion_relation}) which determines the constant $V$.

\begin{center}
\textbf{3.3.4. Equations for ion fluid at the order $\epsilon^{7/6}$}
\end{center}

At the order $\epsilon^{7/6}$, solving the continuity equation and the momentum equation of ion fluid for the unknowns $ n^{(2)} $ and $ u^{(2)} $, we get
\begin{equation}\label{fmKdV_expression_of_n_2}
  n^{(2)} = \frac{\phi^{(2)}}{(V^{2}-\sigma \gamma)}  +  \frac{3V^{2}+\sigma \gamma (\gamma-2)}{2(V^{2}-\sigma \gamma)^{3}} [\phi^{(1)}]^{2} ,
\end{equation}
\begin{equation}\label{fmKdV_expression_of_u_2}
  u^{(2)} = \frac{V\phi^{(2)}}{(V^{2}-\sigma \gamma)}  +  \frac{V(V^{2}+\sigma \gamma^{2})}{2(V^{2}-\sigma \gamma)^{3}} [\phi^{(1)}]^{2} .
\end{equation}

\begin{center}
\textbf{3.3.5. Vlasov-Boltzmann equation at the order $\epsilon^{7/6}$}
\end{center} 

At the order $\epsilon^{7/6}$, the Vlasov-Boltzmann equation for nonthermal electrons is
\begin{equation}\label{mKdV_Vlasov_f_ce31}
 v_{||}\frac{\partial f_{ce}^{(2)}}{\partial \xi} + \frac{\partial \phi^{(2)}}{\partial \xi} \frac{\partial f_{c0}}{\partial v_{||}} + 
 \frac{\partial \phi^{(1)}}{\partial \xi} \frac{\partial f_{ce}^{(1)}}{\partial v_{||}}  = 0.
\end{equation}
Using (\ref{Vlasov_f_ce6}), the equation (\ref{mKdV_Vlasov_f_ce31}) can be simplified as
\begin{equation}\label{mKdV_Vlasov_f_ce31_modified}
 v_{||}\frac{\partial f_{ce}^{(2)}}{\partial \xi} + \frac{\partial \phi^{(2)}}{\partial \xi} \frac{\partial f_{c0}}{\partial v_{||}} + \frac{\partial \psi^{(2)}}{\partial \xi} \frac{\partial g_{c0}}{\partial v_{||}}  = 0 ,
\end{equation}
where
\begin{equation}\label{psi_2_chi_c0}
 \psi^{(2)}=-(\phi^{(1)})^{2} \mbox{     and     } g_{c0} = \frac{\partial f_{c0}}{\partial v_{||}^{2}}.
\end{equation}
If we compare the equation (\ref{mKdV_Vlasov_f_ce31_modified}) with the equation (\ref{fmKdV_Vlasov_f_ce_11}) then we see that the equation (\ref{mKdV_Vlasov_f_ce31_modified}) is exactly similar to the equation (\ref{fmKdV_Vlasov_f_ce_11}) except the presence of the third term in the left hand side of (\ref{mKdV_Vlasov_f_ce31_modified}) and this third term is similar to the second term of the same equation. So we can solve this equation using the same method of solution of the equation (\ref{fmKdV_Vlasov_f_ce_11}) and consequently including an extra higher order time derivative term $\epsilon^{19/6}\alpha_{1}\frac{\partial f_{ce}^{(2)}}{\partial \tau}$, the equation (\ref{mKdV_Vlasov_f_ce31_modified}) can be written in the following form:
\begin{equation}\label{mKdV_Vlasov_f_ce42}
  \alpha_{1} \epsilon^{2}\frac{\partial f_{ce\epsilon}^{(2)}}{\partial \tau} + v_{||}\frac{\partial f_{ce\epsilon}^{(2)}}{\partial \xi} + \frac{\partial \phi^{(2)}}{\partial \xi} \frac{\partial f_{c0}}{\partial v_{||}} + \frac{\partial \psi^{(2)}}{\partial \xi} \frac{\partial g_{c0}}{\partial v_{||}}  = 0,
\end{equation}
where $f_{ce}^{(2)}$ is replaced by $f_{ce\epsilon}^{(2)}$ and $f_{ce}^{(2)}$ can be obtained from the unique solution of equation (\ref{mKdV_Vlasov_f_ce42}) by considering the relation (\ref{f_ce_epsilon_j}) for $j=2$.

As the equation (\ref{mKdV_Vlasov_f_ce42}) is similar to the equation (\ref{fmKdV_Vlasov_f_ce_115}) except the presence of the fourth term of the left hand side of (\ref{mKdV_Vlasov_f_ce42}) which is similar to the third term of the same equation, following the procedure of finding the solution of the equation (\ref{fmKdV_Vlasov_f_ce_115}), the expression of $f_{ce}^{(2)}$ can be written as
\begin{equation}\label{mKdV_IFT_Vlasov_f_ce7}
 f_{ce}^{(2)} = - 2\frac{\partial f_{c0}}{\partial v_{||}^{2}} \phi^{(2)} - 2\frac{\partial g_{c0}}{\partial v_{||}^{2}} \psi^{(2)}.
\end{equation}
Similarly, considering the Vlasov-Boltzmann equation for isothermal electrons at the order $\epsilon^{7/6}$, we get
\begin{equation}\label{mKdV_IFT_Vlasov_f_se7}
 f_{se}^{(2)} = - 2\frac{\partial f_{s0}}{\partial v_{||}^{2}} \phi^{(2)} - 2\frac{\partial g_{s0}}{\partial v_{||}^{2}} \psi^{(2)},
\end{equation}
where 
\begin{equation}\label{g_s0}
 g_{s0} = \frac{\partial f_{s0}}{\partial v_{||}^{2}}.
\end{equation}

\begin{center}
\textbf{3.3.6. Poisson equation at the order $\epsilon^{\frac{2}{3}}$}
\end{center}

From the Poisson equation at the order $\epsilon^{\frac{2}{3}}$, we get
\begin{equation}\label{Poisson_equation_modified_oreder2}
  n^{(2)}  = \int^{\infty}_{-\infty} f_{ce}^{(2)} dv_{\parallel} +\int^{\infty}_{-\infty} f_{se}^{(2)} dv_{||}.
\end{equation}
Substituting (\ref{mKdV_IFT_Vlasov_f_ce7}) and (\ref{mKdV_IFT_Vlasov_f_se7}) into the equation (\ref{Poisson_equation_modified_oreder2}) and using (\ref{fmKdV_expression_of_n_2}), we get
\begin{equation}\label{fmKdV_final_equation_2}
   \Big[ \frac{1}{V^{2}-\sigma \gamma} - (1-\bar{n}_{c0} \sigma_{c}\beta_{e})\Big] \phi^{(2)} + B_{1} [\phi^{(1)}]^{2}  = 0 .
\end{equation}
The above equation is identically satisfied as the first term of the above equation is equal to zero due to the dispersion relation (\ref{dispersion_relation}) and the second term is also equal to zero due to the condition $B_{1}=0$. Therefore, Poisson equation at the order $\epsilon^{\frac{2}{3}}$ is identically satisfied if the condition $B_{1}=0$ holds good.

\begin{center}
\textbf{3.3.7. Equations for ion fluid at the order $\epsilon^{9/6}$}
\end{center}

Again, at the order $\epsilon^{9/6}$, solving the continuity equation of ions and the momentum equation of ions for the unknowns $n^{(3)}$ and $u^{(3)}$, we obtain the following equations:
\begin{eqnarray}\label{fmKdV_expression_of_n_3}
   n^{(3)} = \frac{\phi^{(3)}}{V^{2}-\sigma \gamma}  + \frac{3V^{2} + \sigma \gamma (\gamma-2)}{(V^{2}-\sigma \gamma)^{3}} \phi^{(1)}\phi^{(2)}+ \frac{1}{6} H_{2} [\phi^{(1)}]^{3} ,
\end{eqnarray}
\begin{eqnarray}\label{fmKdV_expression_of_u_3}
  u^{(3)} = \frac{V \phi^{(3)}}{V^{2}-\sigma \gamma}  +  \frac{V(V^{2}+\sigma \gamma^{2})}{(V^{2}-\sigma \gamma)^{3}} \phi^{(1)}\phi^{(2)} + G_{2} [\phi^{(1)}]^{3}, 
\end{eqnarray}
where
\begin{equation}\label{G_2}
  G_{2}= \frac{V}{6} \Big[\frac{3V^{4} + \sigma \gamma^{2} (\gamma+7)V^{2} + \sigma^{2} \gamma^{3} (2\gamma-1)}{(V^{2}-\sigma \gamma)^{5}} \Big].  
\end{equation}

\begin{center}
\textbf{3.3.8. Vlasov-Boltzmann equation at the order $\epsilon^{9/6}$}
\end{center} 

At the order $\epsilon^{9/6}$, the Vlasov-Boltzmann equation of nonthermal electrons (\ref{f_ce}) is
\begin{equation}\label{fmKdV_Vlasov_f_ce_31}
  v_{||}\frac{\partial f_{ce}^{(3)}}{\partial \xi} + \frac{\partial \phi^{(3)}}{\partial \xi} \frac{\partial f_{c0}}{\partial v_{||}}  
  + \frac{\partial \phi^{(2)}}{\partial \xi} \frac{\partial f_{ce}^{(1)}}{\partial v_{||}} + \frac{\partial \phi^{(1)}}{\partial \xi} \frac{\partial f_{ce}^{(2)}}{\partial v_{||}} = 0 .
\end{equation}
Using the equations (\ref{Vlasov_f_ce6}) and (\ref{mKdV_IFT_Vlasov_f_ce7}), the equation (\ref{fmKdV_Vlasov_f_ce_31}) can be simplified as
\begin{equation}\label{fmKdV_Vlasov_f_ce31_modified}
 v_{||}\frac{\partial f_{ce}^{(3)}}{\partial \xi} + \frac{\partial \phi^{(3)}}{\partial \xi} \frac{\partial f_{c0}}{\partial v_{||}} + \frac{\partial \psi^{(3)}}{\partial \xi} \frac{\partial g_{c0}}{\partial v_{||}}+ \frac{\partial \chi^{(3)}}{\partial \xi} \frac{\partial h_{c0}}{\partial v_{||}}  = 0 ,
\end{equation}
where
\begin{equation}\label{psi_3_chi_3}
 \psi^{(3)}=-2\phi^{(1)}\phi^{(2)},\chi^{(3)}=\frac{2}{3}(\phi^{(1)})^{3},h_{c0} = \frac{\partial g_{c0}}{\partial v_{||}^{2}}.
\end{equation}
The equation (\ref{fmKdV_Vlasov_f_ce31_modified}) is qualitatively similar to the equation (\ref{fmKdV_Vlasov_f_ce_11}) except the presence of the third and the fourth term of the left hand side of (\ref{fmKdV_Vlasov_f_ce31_modified}). These two terms are qualitatively similar to the second term of the same equation. So using the same method of finding the solution of the equation (\ref{fmKdV_Vlasov_f_ce_11}), we add an extra higher order time derivative term $\epsilon^{21/6}\alpha_{1}\frac{\partial f_{ce}^{(3)}}{\partial \tau}$ with the Vlasov-Boltzmann equation at the order $\epsilon^{9/6}$ and the unique solution of the equation (\ref{fmKdV_Vlasov_f_ce31_modified}) can be written in the following form:
\begin{equation}\label{fmKdV_Vlasov_f_ce7}
 f_{ce}^{(3)} = - 2\frac{\partial f_{c0}}{\partial v_{||}^{2}} \phi^{(3)} - 2\frac{\partial g_{c0}}{\partial v_{||}^{2}} \psi^{(3)}- 2\frac{\partial h_{c0}}{\partial v_{||}^{2}} \chi^{(3)}.
\end{equation}
Similarly, considering the Vlasov-Boltzmann equation of isothermal electrons at the order $\epsilon^{9/6}$, we get
\begin{equation}\label{fmKdV_Vlasov_f_se7}
 f_{se}^{(3)} = - 2\frac{\partial f_{s0}}{\partial v_{||}^{2}} \phi^{(3)} - 2\frac{\partial g_{s0}}{\partial v_{||}^{2}} \psi^{(3)}- 2\frac{\partial h_{s0}}{\partial v_{||}^{2}} \chi^{(3)},
\end{equation}
where
\begin{equation}\label{h_s0}
 h_{s0}=\frac{\partial g_{s0}}{\partial v_{||}^{2}}.
\end{equation}

\begin{center}
\textbf{3.3.9. Poisson equation at the order $\epsilon$}
\end{center} 

From the Poisson equation at the order $\epsilon$, we get
\begin{eqnarray}\label{Poisson_equation_modified_oreder3}
  n^{(3)}  = \int^{\infty}_{-\infty} f_{ce}^{(3)} dv_{\parallel} +\int^{\infty}_{-\infty} f_{se}^{(3)} dv_{||}.
\end{eqnarray}
Using (\ref{fmKdV_Vlasov_f_ce7}), (\ref{fmKdV_Vlasov_f_se7}) and (\ref{fmKdV_expression_of_n_3}), the Poisson equation (\ref{Poisson_equation_modified_oreder3}) can be written as 
\begin{eqnarray}\label{fmKdV_final_equation_3}
	 \bigg[ \frac{1}{(V^{2}-\sigma \gamma)} - (1-\bar{n}_{c0} \sigma_{c}\beta_{e})\bigg] \phi^{(3)} + 2B_{1} [\phi^{(1)}\phi^{(2)}] + \frac{2}{3} B_{2} [\phi^{(1)}]^{3} = 0.
\end{eqnarray}
The above equation is identically satisfied as the first term is equal to zero due to the dispersion relation (\ref{dispersion_relation}) whereas the second and third terms can be made equal to zero if we use the conditions $B_{1}=0$ and $B_{2}=0$ respectively. Therefore, at the order $\epsilon$, the Poisson equation is also identically satisfied if $B_{1}=0$ and $B_{2}=0$.

\begin{center}
\textbf{3.3.10. Equations for ion fluid at the order $\epsilon^{11/6}$} 
\end{center}  

At the order $\epsilon^{11/6}$, solving the continuity equation and the momentum equation of ions, $ \frac{\partial n^{(4)}}{\partial \xi} $ and $ \frac{\partial u^{(4)}}{\partial \xi} $ can be expressed as functions of $\phi^{(1)}$, $\phi^{(2)}$, $\phi^{(3)}$ and $\phi^{(4)}$ along with their different derivatives with respect to $\xi$ and $\tau$. In particular, $ \frac{\partial n^{(4)}}{\partial \xi} $ can be written as
\begin{eqnarray}\label{fmKdV_expression_of_n_4}
   \frac{\partial n^{(4)}}{\partial \xi} = \frac{1}{(V^{2}-\sigma \gamma)} \frac{\partial \phi^{(4)}}{\partial \xi} + \frac{2V}{(V^{2}-\sigma \gamma)^{2}} \frac{\partial \phi^{(1)}}{\partial \tau} \nonumber \\
  + \frac{3V^{2}+\sigma \gamma (\gamma-2)}{(V^{2}-\sigma \gamma)^{3}} \frac{\partial}{\partial \xi}[\phi^{(1)}\phi^{(3)}+\frac{1}{2}(\phi^{(2)})^{2}] \nonumber \\ + \frac{H_{2}}{2}  \frac{\partial}{\partial \xi}[(\phi^{(1)})^{2}\phi^{(2)}]  
	   + \frac{H_{3}}{6}  [\phi^{(1)}]^{3} \frac{\partial \phi^{(1)}}{\partial \xi} ,
\end{eqnarray}
where
\begin{eqnarray}\label{fmKdV_expression_of_H_3}
   H_{3} = \frac{1}{(V^{2}-\sigma \gamma)^{7}} \Big[105V^{6}
   + \sigma \gamma (\gamma^{3}+21\gamma^{2}+161\gamma-174)V^{4} \nonumber \\
	+ \sigma^{2} \gamma^{2} (8\gamma^{3}+53\gamma^{2}-162\gamma+108)V^{2}\nonumber \\
	+ \sigma^{3} \gamma^{3} (\gamma-2)(2\gamma-3)(3\gamma-4) \Big] .
\end{eqnarray}

\begin{center}
\textbf{3.3.11. Vlasov-Boltzmann equation at the order $\epsilon^{11/6}$}
\end{center}

At the order $\epsilon^{11/6}$, the Vlasov-Boltzmann equation of nonthermal electrons is
\begin{eqnarray}\label{fmKdV_Vlasov_f_ce_41_modified}
v_{||}\frac{\partial f_{ce}^{(4)}}{\partial \xi} + \frac{\partial \phi^{(4)}}{\partial \xi} \frac{\partial f_{c0}}{\partial v_{||}} + \frac{\partial \psi^{(4)}}{\partial \xi} \frac{\partial g_{c0}}{\partial v_{||}} + \frac{\partial \chi^{(4)}}{\partial \xi} \frac{\partial h_{c0}}{\partial v_{||}}+ \frac{\partial \kappa^{(4)}}{\partial \xi} \frac{\partial k_{c0}}{\partial v_{||}}  + 2 
 \alpha_{1} V \frac{\partial f_{c0}}{\partial v_{||}^{2}}\frac{\partial \phi^{(1)}}{\partial \xi}   = 0 ,
\end{eqnarray}
where we have used the equations (\ref{Vlasov_f_ce6}), (\ref{mKdV_IFT_Vlasov_f_ce7}) and (\ref{fmKdV_Vlasov_f_ce7}) to get the equation (\ref{fmKdV_Vlasov_f_ce_41_modified}) and in this equation, we have used the following notations:
\begin{equation}\label{psi_4_chi_4_etc1}
 \psi^{(4)}=-2\phi^{(1)}\phi^{(3)}-(\phi^{(2)})^{2},
 \chi^{(4)}=2(\phi^{(1)})^{2}\phi^{(2)},
\kappa^{(4)}=-\frac{1}{3}(\phi^{(1)})^{4}, k_{c0} = \frac{\partial h_{c0}}{\partial v_{||}^{2}}.\nonumber
\end{equation}
If we ignore the last term of the left hand side of the equation (\ref{fmKdV_Vlasov_f_ce_41_modified}), this equation is qualitatively similar to the equation (\ref{fmKdV_Vlasov_f_ce_11}) and we can easily find the solution of the equation (\ref{fmKdV_Vlasov_f_ce_41_modified}). In fact, the equation (\ref{fmKdV_Vlasov_f_ce_41_modified}) is complicated one because of the presence of the last term. It is also evident to note that the last term of the left hand side of the equation (\ref{fmKdV_Vlasov_f_ce_41_modified}) is qualitatively different from the second or third or fourth or fifth term of the left hand side of the equation (\ref{fmKdV_Vlasov_f_ce_41_modified}) because $\displaystyle \frac{\partial f_{c0}}{\partial v_{||}^{2}}$ is a function of $v_{||}^{2}$ whereas $\displaystyle \frac{\partial f_{c0}}{\partial v_{||}} $, $\displaystyle \frac{\partial g_{c0}}{\partial v_{||}} $, $\displaystyle \frac{\partial h_{c0}}{\partial v_{||}}$, $\displaystyle \frac{\partial k_{c0}}{\partial v_{||}}$ are not functions of $v_{||}^{2}$. In particular, it is simple to note
\begin{equation}\label{simple}
\frac{\partial }{\partial v_{||}} (f_{c0},g_{c0},h_{c0},k_{c0})=2v_{||}\frac{\partial }{\partial v_{||}^{2}} (f_{c0},g_{c0},h_{c0},k_{c0}).
\end{equation}
Presence of $v_{||}$ in the numerator of the above mentioned terms makes the algebra very simple and consequently one can easily get the solution of the lower order Vlasov-Boltzmann equations. So, including an extra higher order time derivative term $\epsilon^{23/6}\alpha_{1}\frac{\partial f_{ce}^{(4)}}{\partial \tau}$, the equation (\ref{fmKdV_Vlasov_f_ce_41_modified}) can be written as
\begin{eqnarray}\label{fmKdV_Vlasov_f_ce_41_modified_modified}
\alpha_{1} \epsilon^{2} \frac{\partial f_{ce\epsilon}^{(4)}}{\partial \tau}+v_{||}\frac{\partial f_{ce\epsilon}^{(4)}}{\partial \xi} + \frac{\partial \phi^{(4)}}{\partial \xi} \frac{\partial f_{c0}}{\partial v_{||}} + \frac{\partial \psi^{(4)}}{\partial \xi} \frac{\partial g_{c0}}{\partial v_{||}}+ \frac{\partial \chi^{(4)}}{\partial \xi} \frac{\partial h_{c0}}{\partial v_{||}} + \frac{\partial \kappa^{(4)}}{\partial \xi} \frac{\partial k_{c0}}{\partial v_{||}}\nonumber \\ + 2 \alpha_{1} V \frac{\partial f_{c0}}{\partial v_{||}^{2}}\frac{\partial \phi^{(1)}}{\partial \xi}   = 0 ,
\end{eqnarray}
where $f_{ce}^{(4)}$ is replaced by $f_{ce\epsilon}^{(4)}$ and $f_{ce}^{(4)}$ can be obtained from the unique solution of equation (\ref{fmKdV_Vlasov_f_ce_41_modified_modified}) by considering the relation (\ref{f_ce_epsilon_j}) for $j=4$.

Now, assuming the $\tau$ dependence of the perturbed quantities is of the form $\exp(i\omega \tau)$ and taking the Fourier transform with respect to $\xi$, we get the following equation from the equation (\ref{fmKdV_Vlasov_f_ce_41_modified_modified}):
\begin{eqnarray}\label{fmKdV_Vlasov_f_ce_41_modified_modified_modified}
i\widetilde{f}_{ce\epsilon}^{(4)} = -\frac{2}{s}\Big[\frac{\partial f_{c0}}{\partial v_{||}^{2}}\widetilde{\phi}_{\xi}^{(4)} + \frac{\partial g_{c0}}{\partial v_{||}^{2}}\widetilde{\psi}_{\xi}^{(4)} +\frac{\partial h_{c0}}{\partial v_{||}^{2}}\widetilde{\chi}_{\xi}^{(4)} + \frac{\partial k_{c0}}{\partial v_{||}^{2}}\widetilde{\kappa}_{\xi}^{(4)}\Big] \times\frac{sv_{||}}{sv_{||}+\alpha_{1} \omega \epsilon^{2}}\nonumber \\- 2 \alpha_{1} V \frac{\partial f_{c0}}{\partial v_{||}^{2}}\frac{1}{sv_{||}+\alpha_{1} \omega \epsilon^{2}}\widetilde{\phi}_{\xi}^{(1)} ,
\end{eqnarray}
where $\widetilde{\phi}_{\xi}^{(4)}$, $\widetilde{\psi}_{\xi}^{(4)}$, $\widetilde{\chi}_{\xi}^{(4)}$, $\widetilde{\kappa}_{\xi}^{(4)}$ and $\widetilde{\phi}_{\xi}^{(1)}$ are, respectively, the Fourier transform of $\phi_{\xi}^{(4)}$, $\psi_{\xi}^{(4)}$, $\chi_{\xi}^{(4)}$, $\kappa_{\xi}^{(4)}$ and $\phi_{\xi}^{(1)}$.
  
Now, making $\epsilon \to 0$ and using the relations $x\mathcal{P}(1/x)=1$, $x\delta(x)=0$ and $s\delta(sv_ {||})=\mbox{sgn}(s)\delta(v_ {||})$, we get the following expression of $\widetilde{f}_{ce}^{(4)}$:
\begin{eqnarray}\label{Vlasov_f_ce7}
 is\widetilde{f}_{ce}^{(4)} = -2\Big[\frac{\partial f_{c0}}{\partial v_{||}^{2}}\widetilde{\phi}_{\xi}^{(4)} + \frac{\partial g_{c0}}{\partial v_{||}^{2}}\widetilde{\psi}_{\xi}^{(4)} +\frac{\partial h_{c0}}{\partial v_{||}^{2}}\widetilde{\chi}_{\xi}^{(4)} + \frac{\partial k_{c0}}{\partial v_{||}^{2}}\widetilde{\kappa}_{\xi}^{(4)}\Big]\nonumber \\ - 2 
 \alpha_{1} V \frac{\partial f_{c0}}{\partial v_{||}^{2}}  \Big[sP\Big(\frac{1}{sv_{||}}\Big) + i\pi\mbox{sgn}(s)\delta(v_ {||})\Big]\widetilde{\phi}_{\xi}^{(1)} .
\end{eqnarray}
Integrating (\ref{Vlasov_f_ce7}) over the velocity space, we get
\begin{eqnarray}\label{Vlasov_f_ce7_modified}
is\int_{-\infty}^{+\infty}\widetilde{f}_{ce}^{(4)}dv_{||} = -2\Big[F_{c0}\widetilde{\phi}_{\xi}^{(4)} + G_{c0}\widetilde{\psi}_{\xi}^{(4)} +H_{c0}\widetilde{\chi}_{\xi}^{(4)} + K_{c0}\widetilde{\kappa}_{\xi}^{(4)}\Big]\nonumber \\- 2i \pi \alpha_{1} V Z_{c0}\mbox{sgn}(s)\widetilde{\phi}_{\xi}^{(1)}, 
\end{eqnarray}
where $F_{c0}$, $G_{c0}$, $H_{c0}$, $K_{c0}$, $Z_{c0}$ are given in Appendix A.

Taking Fourier inversion of (\ref{Vlasov_f_ce7_modified}), we get
\begin{eqnarray}\label{Vlasov_f_ce7_further_modified}
\frac{\partial}{\partial \xi}\Bigg(\int_{-\infty}^{+\infty}f_{ce}^{(4)}dv_{||}\Bigg) = -2\Big[F_{c0}\phi_{\xi}^{(4)} + G_{c0}\psi_{\xi}^{(4)} +H_{c0}\chi_{\xi}^{(4)} + K_{c0}\kappa_{\xi}^{(4)}\Big]\nonumber \\ + 2 \alpha_{1} V Z_{c0}\mathcal{P} \int^{\infty}_{-\infty} \frac{\partial \phi^{(1)}}{\partial \xi'} \frac{d \xi'}{\xi-\xi'} ,
\end{eqnarray}
where we have used the Convolution theorem of Fourier transform to find the inverse Fourier transform of $\displaystyle \mbox{sgn}(s)\widetilde{\phi}_{\xi}^{(1)}$. Here $\displaystyle \frac{\partial \phi^{(1)}}{\partial \xi'}$ is the value of $\displaystyle \frac{\partial \phi^{(1)}}{\partial \xi}$ at $\xi=\xi'$.

Similarly, considering the Vlasov-Boltzmann equation of the isothermal electrons at the order $\epsilon^{11/6}$, we get
\begin{eqnarray}\label{Vlasov_f_se7}
\frac{\partial}{\partial \xi}\Bigg(\int_{-\infty}^{+\infty}f_{se}^{(4)}dv_{||}\Bigg)= -2\Big[F_{s0}\phi_{\xi}^{(4)} + G_{s0}\psi_{\xi}^{(4)} +H_{s0}\chi_{\xi}^{(4)} + K_{s0}\kappa_{\xi}^{(4)}\Big]\nonumber \\ + 2 \alpha_{1} V Z_{s0}\mathcal{P} \int^{\infty}_{-\infty} \frac{\partial \phi^{(1)}}{\partial \xi'} \frac{d \xi'}{\xi-\xi'} ,
\end{eqnarray}
where $F_{s0}$, $G_{s0}$, $H_{s0}$, $K_{s0}$, $Z_{s0}$ are given in Appendix B.

\begin{center}
\textbf{3.3.12. Poisson equation at the order $\epsilon^{4/3}$}
\end{center} 

From the Poisson equation at the order $\epsilon^{\frac{4}{3}}$, we get
\begin{eqnarray}\label{fmKdV_poisson_equation_4}
  n^{(4)} = \int^{\infty}_{-\infty} f_{ce}^{(4)} dv_{\parallel}  + \int^{\infty}_{-\infty} f_{se}^{(4)} dv_{||}  -  \frac{\partial^{2} \phi^{(1)}}{\partial \xi^{2}} .
\end{eqnarray}
Differentiating this equation with respect to $\xi$, using the equations  (\ref{Vlasov_f_ce7_further_modified}) and (\ref{Vlasov_f_se7}) in the resulting equation, we get the following expression of $\displaystyle \frac{\partial n^{(4)}}{\partial \xi}$ as follows:
\begin{eqnarray}\label{fmKdV_expression_of_n_4_Vlasov}
  \frac{\partial n^{(4)}}{\partial \xi}  &=&  - \frac{\partial^{3} \phi^{(1)}}{\partial \xi^3}+ (1-\bar{n}_{c0} \sigma_{c}\beta_{e}) \frac{\partial \phi^{(4)}}{\partial \xi}\nonumber \\
	  &+& \big(\bar{n}_{c0} \sigma_{c}^2 + \bar{n}_{s0} \sigma_{s}^2\big) \frac{\partial}{\partial \xi}[\phi^{(1)}\phi^{(3)}+\frac{1}{2}(\phi^{(2)})^{2}] \nonumber\\
	&+& \frac{1}{2}\Big[\bar{n}_{c0} \sigma_{c}^3 (1+3\beta_{e}) + \bar{n}_{s0} \sigma_{s}^3 \Big] \frac{\partial}{\partial \xi}[(\phi^{(1)})^{2}\phi^{(2)}] \nonumber \\
&+& \frac{1}{6}\Big[\bar{n}_{c0} \sigma_{c}^4 (1+8\beta_{e}) + \bar{n}_{s0} \sigma_{s}^4 \Big] [\phi^{(1)}]^{3} \frac{\partial \phi^{(1)}}{\partial \xi}
	  \nonumber \\
&-& \frac{\alpha_{1}V}{\sqrt{2\pi}} \bigg[\bar{n}_{c0} \sigma_{c}^{\frac{3}{2}}(1-\frac{3}{4}\beta_{e}) + \bar{n}_{s0} \sigma_{s}^{\frac{3}{2}}\bigg] \mathcal{P} \int^{\infty}_{-\infty} \frac{\partial \phi^{(1)}}{\partial \xi'} \frac{d \xi'}{\xi-\xi'} .
\end{eqnarray}

Now, eliminating $\frac{\partial n^{(4)}}{\partial \xi}$ from equations (\ref{fmKdV_expression_of_n_4}) and (\ref{fmKdV_expression_of_n_4_Vlasov}), we get
\begin{eqnarray}\label{FmKdV_KdV_equation}
  \frac{\partial \phi^{(1)}}{\partial \tau} + AB_{3} [\phi^{(1)}]^{3} \frac{\partial \phi^{(1)}}{\partial \xi} + \frac{1}{2}A \frac{\partial^{3} \phi^{(1)}}{\partial \xi^3} + \frac{1}{2}AE \alpha_{1} \mathcal{P} \int^{\infty}_{-\infty} \frac{\partial \phi^{(1)}}{\partial \xi'} \frac{d \xi'}{\xi-\xi'} = 0 ,
\end{eqnarray}
where we have used the dispersion relation (\ref{dispersion_relation}), conditions $B_{1}=0$ and $B_{2}=0$ to eliminate  the terms $\displaystyle \frac{\partial \phi^{(4)}}{\partial \xi}$, $\displaystyle AB_{1}\frac{\partial}{\partial \xi}[\phi^{(1)}\phi^{(3)}+\frac{1}{2}(\phi^{(2)})^{2}]$ and $\displaystyle AB_{2} \frac{\partial}{\partial \xi}[(\phi^{(1)})^{2}\phi^{(2)}]$ respectively to simplify the equation (\ref{FmKdV_KdV_equation}).

\noindent Here, $B_{3}$ is given by
\begin{equation}\label{fmKdV_B_3}
		B_{3} = \frac{1}{12}  \Big[ H_{3} - \Big(\bar{n}_{c0} \sigma_{c}^4 (1+8\beta_{e}) + \bar{n}_{s0} \sigma_{s}^4 \Big) \Big] ,
\end{equation}
where $H_{3}$ is given by the equation (\ref{fmKdV_expression_of_H_3}).

Therefore, the Poisson equation at the order $\epsilon^{\frac{4}{3}}$ gives a FMKdV equation including the effect of Landau damping which describes the nonlinear behaviour of IA waves when $B_{1}=0$, $B_{2}=0$ but $B_{3} \neq 0$.

\begin{center}
\textbf{4.  SOLITARY WAVE SOLUTION}
\end{center} 

In more compact form, we can write the equations (\ref{KdV_equation}), (\ref{mKdV_KdV_equation}) and (\ref{FmKdV_KdV_equation}) as a single equation:
\begin{eqnarray}\label{General_KdV_equation}
   \frac{\partial \phi^{(1)}}{\partial \tau} + AB_{r} [\phi^{(1)}]^{r} \frac{\partial \phi^{(1)}}{\partial \xi} + \frac{1}{2}A \frac{\partial^{3} \phi^{(1)}}{\partial \xi^3}+ \frac{1}{2}AE \alpha_{1} \mathcal{P} \int^{\infty}_{-\infty} \frac{\partial \phi^{(1)}}{\partial \xi'} \frac{d \xi'}{\xi-\xi'} = 0 ,
\end{eqnarray}
where $r=1, 2, 3$. 

In this section, we have investigated the solitary wave solution of the equation (\ref{General_KdV_equation}). If we put $\alpha_{1}=0$ in equation (\ref{General_KdV_equation}), then the equation (\ref{General_KdV_equation}) reduces to a KdV equation for $r=1$, a MKdV equation for $r=2$ and a FMKdV equation for $r=3$.

For a solitary wave solution of (\ref{General_KdV_equation}) with $\alpha_{1}=0$, we consider the following transformation of the independent variables:
\begin{equation}\label{Transformation_solution}
  X=\xi-U\tau , \tau'=\tau .
\end{equation}
Under the above transformation of independent variables, the equation (\ref{General_KdV_equation}) with $\alpha_{1}=0$ assumes the following form:
\begin{equation}\label{Solution_KdV_equation}
  \frac{\partial \phi^{(1)}}{\partial \tau} - U \frac{\partial \phi^{(1)}}{\partial X} + AB_{r} [\phi^{(1)}]^{r} \frac{\partial \phi^{(1)}}{\partial X} + \frac{1}{2}A \frac{\partial^{3} \phi^{(1)}}{\partial X^3} = 0 ,
\end{equation}
where we drop the prime on the independent variable $\tau$ to simplify the notation.
 
For the travelling wave solution of (\ref{Solution_KdV_equation}), we take 
\begin{equation}\label{Solution_phi_0}
  \phi^{(1)}= \phi_{0}(X) .
\end{equation}

Substituting (\ref{Solution_phi_0}) into (\ref{Solution_KdV_equation}), we get an ordinary differential equation of $\phi_{0}$ with respect to the independent variable $X$ and the solitary wave solution of the resulting differential equation of $\phi_{0}$ can be written as
\begin{equation}\label{Solution_phi_0_1}
  \phi_{0} = a ~  \mbox{sech}^{\frac{2}{r}} \big[ WX \big] ,
\end{equation}
where we have used the boundary conditions $\phi_{0}$, $\frac{d\phi_{0}}{dX}$, $\frac{d^{2}\phi_{0}}{dX^{2}}$, $\frac{d^{3}\phi_{0}}{dX^{3}}$ $\rightarrow 0$ as $|X|\rightarrow \infty$. Here, the amplitude ($a$) and width ($\frac{1}{W}$) are given by
\begin{equation}\label{Solution_a_W}
  a^{r} = \frac{(r+1)(r+2)U}{2AB_{r}} ~ \mbox{  and  } ~ W^{2} = \frac{r^{2}U}{2A} .
\end{equation}
Now, using (\ref{Solution_a_W}), the equation (\ref{Solution_phi_0_1}) can be written as
\begin{equation}\label{Solution_phi_0_new}
  \phi_{0} = a~\mbox{sech}^{\frac{2}{r}} \Big[\sqrt{\frac{r^{2}a^{r}B_{r}}{(r+1)(r+2)}} \Big\{\xi-\frac{2AB_{r}a^{r} \tau}{(r+1)(r+2)}\Big\} \Big] .
\end{equation}

From (\ref{General_KdV_equation}), we have 
\begin{eqnarray}\label{Ott_and_Sudan_condition}
  \frac{\partial}{\partial \tau}  \int_{-\infty}^{\infty}\Big(\phi^{(1)}\Big)^{2}d\xi = -AE\alpha_{1} \int^{\infty}_{-\infty} \phi^{(1)} \Big[ \mathcal{P} \int^{\infty}_{-\infty} \frac{\partial \phi^{(1)}}{\partial \xi'} \frac{d \xi'}{\xi-\xi'}\Big]d\xi,
\end{eqnarray}
where we have used the boundary conditions for the solitary wave solution, i.e.,  $\phi^{(1)}$, $\frac{\partial\phi^{(1)}}{\partial\xi}$, $\frac{\partial^{2}\phi^{(1)}}{\partial\xi^{2}}$ $\rightarrow 0$ as $|\xi|\rightarrow \infty$. If the initial perturbation is of the form (\ref{Solution_phi_0_new}), then one can check that the integral appearing in the right hand side of (\ref{Ott_and_Sudan_condition}) is positive for $r=1,2,3$ and consequently from (\ref{Ott_and_Sudan_condition}), we have 
\begin{eqnarray}\label{Ott_and_Sudan_condition_simplified}
  \frac{\partial}{\partial \tau}  \int_{-\infty}^{\infty}\Big(\phi^{(1)}\Big)^{2}d\xi <0,
\end{eqnarray}
for any values of the parameters of the system because $A$, $E$, $\alpha_{1}$ are all strictly positive. The inequality (\ref{Ott_and_Sudan_condition_simplified}) shows that the initial perturbation of the form (\ref{Solution_phi_0_new}) will decay to zero. This phenomenon suggests that the wave amplitude is not a constant but decreases slowly with time. 

Now, using the prescription of Ott and Sudan [45], we have introduced the following space coordinate:
\begin{equation}\label{Expression_of_X}
  X = \sqrt{\frac{r^{2}a^{r}B_{r}}{(r+1)(r+2)}} \Big[\xi-\frac{2AB_{r}}{(r+1)(r+2)} \int_{0}^{\tau} a^{r} d\tau\Big] ,
\end{equation}
where the amplitude ($a$) is a slowly varying function of time. Again, we consider $\phi^{(1)}$ as a function of $X$ and $\tau$, i.e., $\phi^{(1)}=\phi^{(1)}(X,\tau)$ and consequently the equation (\ref{General_KdV_equation}) assumes the following form:
\begin{eqnarray}\label{Solution_General_KdV_equation}
   \frac{\partial \phi^{(1)}}{\partial \tau} 
	    + \Big[ - \frac{2AB_{r}a^{r}W}{(r+1)(r+2)} + \frac{rX}{2a} \frac{\partial a}{\partial \tau} \Big] \frac{\partial \phi^{(1)}}{\partial X} + AB_{r}W \Big(\phi^{(1)}\Big)^{r} \frac{\partial \phi^{(1)}}{\partial X}  + \frac{1}{2}AW^{3} \frac{\partial^{3} \phi^{(1)}}{\partial X^3} \nonumber \\ + \frac{1}{2}AEW \alpha_{1} \mathcal{P} \int^{\infty}_{-\infty} \frac{\partial \phi^{(1)}}{\partial X'} \frac{d X'}{X-X'} = 0 ,
\end{eqnarray}
where $\displaystyle \frac{\partial \phi^{(1)}}{\partial X'} = \frac{\partial \phi^{(1)}}{\partial X}$ at $X=X'$.

According to multiple time scale analysis [45,48,49] with respect to the small parameter $\alpha_{1}$, we take
\begin{equation}\label{Solution_Expression_of_phi_1}
  \phi^{(1)}(X,\tau) = \sum_{j=0}^{\infty} \alpha_{1}^{j} q^{(j)} ,
\end{equation}
where $q^{(j)}=q^{(j)}(X,\tau_{0}, \tau_{1}, \tau_{2},\cdots)$ for $j=0,1,2,3,\cdots$ and $\tau_{j}$ is given by
\begin{equation}\label{Solution_Expression_of_phi_2}
    \tau_{j}= \alpha_{1}^{j}\tau , j=0,1,2,3,\cdots  .
\end{equation}

Substituting (\ref{Solution_Expression_of_phi_1}) into (\ref{Solution_General_KdV_equation}) and equating the coefficients of different powers of $\alpha_{1}$ on each side of the resulting equation, we get a sequence of equations. The zeroth and the first order equations are, respectively, given by the following equations:
\begin{equation}\label{Solution_Lagrange_equation_q_0}
  \rho \bigg[ \frac{\partial q^{(0)}}{\partial \tau_{0}} + \frac{rX}{2a} \frac{\partial a}{\partial \tau_{0}} \frac{\partial q^{(0)}}{\partial X} \bigg] +  L\Big[\frac{\partial q^{(0)}}{\partial X}\Big]=0  ,
\end{equation}
\begin{equation}\label{Solution_Lagrange_equation_q_1}
  \rho \bigg[ \frac{\partial q^{(1)}}{\partial \tau_{0}} + \frac{rX}{2a} \frac{\partial a}{\partial \tau_{0}} \frac{\partial q^{(1)}}{\partial X} \bigg]  + \frac{\partial [Lq^{(1)}] }{\partial X}  = \rho M q^{(0)} ,
\end{equation}
where 
\begin{equation}\label{Solution_rho}
  \rho=\frac{2}{AW^{3}}, 
\end{equation}
\begin{equation}\label{Solution_L_operator}
  L = \frac{\partial^{2}}{\partial X^{2}}  + \frac{2(r+1)(r+2)}{r^{2}a^{r}} (q^{(0)})^{r} -\frac{4}{r^2} ,
\end{equation}
\begin{eqnarray}\label{Solution_M_q_0_operator}
  -M q^{(0)} = \frac{\partial q^{(0)}}{\partial \tau_{1}} + \frac{rX}{2a} \frac{\partial a}{\partial \tau_{1}}\frac{\partial q^{(0)}}{\partial X}  +  \frac{1}{2}AEW \mathcal{P} \int^{\infty}_{-\infty} \frac{\partial q^{(0)}}{\partial X'} \frac{d X'}{X-X'} .
\end{eqnarray}

Now, it is simple to check that  $q^{(0)}=a ~ \mbox{sech}^{\frac{2}{r}} [X] $ is the soliton solution of (\ref{Solution_Lagrange_equation_q_0}) if
and only if
\begin{equation}\label{condition_on_tau_0}
  \frac{\partial a}{\partial \tau_{0}}=0.
\end{equation}
Using (\ref{condition_on_tau_0}), the equation (\ref{Solution_Lagrange_equation_q_1}) can be written as 
\begin{equation}\label{Solution_Lagrange_equation_q_1_reduced}
  \rho \frac{\partial q^{(1)} }{\partial \tau_{0}}  + \frac{\partial }{\partial X} [Lq^{(1)}] = \rho M q^{(0)} .
\end{equation}
Now, for the existence of the solution of equation (\ref{Solution_Lagrange_equation_q_1_reduced}), we have the following consistency condition: 
\begin{equation}\label{Solution_Kernel}
   \int_{-\infty}^{\infty} \mbox{sech}^{\frac{2}{r}} [X] M q^{(0)} dX =0 .
\end{equation}
The above equation states that the right hand side of (\ref{Solution_Lagrange_equation_q_1_reduced}) is perpendicular to the kernel of adjoint operator of $\frac{\partial }{\partial X} [L]$ and this kernel is $\mbox{sech}^{\frac{2}{r}} [X]$, which tends to zero as $|X| \rightarrow \infty$.  
The equation (\ref{Solution_Kernel}) gives the following differential equation for the solitary wave amplitude $a$:
\begin{equation}\label{Solution_diff_eqn_amplitude}
   M_{r} \frac{\partial }{\partial \tau_{1}}\Big(\frac{a}{a_{0}}\Big) + W_{r} \Big(\frac{a}{a_{0}}\Big)^{\frac{r}{2}+1} =0 ,
\end{equation}
where $a_{0}$ is the value of $a$ when $\tau=0$ and
\begin{equation}\label{Solution_expression_M_r}
   M_{r} = \int_{-\infty}^{\infty} [\mbox{sech \textit{X}} ]^{\frac{4}{r}} (1-X \tanh X)dX,
\end{equation}
\begin{equation}\label{Solution_expression_W_r}
    W_{r}  = \frac{1}{2} I_{r} A E \sqrt{\frac{r^{2}a_{0}^{r}B_{r}}{(r+1)(r+2)}},
\end{equation}
\begin{equation}\label{Solution_expression_I_r}
    I_{r}  = \mathcal{P} \int \limits_{-\infty}^{+\infty}\int \limits_{-\infty}^{+\infty} [\mbox{sech \textit{X}} ]^{\frac{2}{r}} \frac{\partial [\mbox{sech \textit{X}}' ]^{\frac{2}{r}}}{\partial X'}  \frac{dX dX'}{X-X'}   .
\end{equation}

Now, it is simple to check that $M_{1}=1$, $M_{2}=1$, $M_{3} \approx 0.6468$. In Appendix - C, we have generalised the method of Weiland et al. [55] to find $I_{r}$. Using this method and MATHEMATICA [56], we get the following numerical values of $I_{r}$ for $r=1,2,3$ : $I_{1} \approx 2.9231$, $I_{2} \approx 2.7726$, $I_{3} \approx 2.6649$. 

For $r=1,2,3$ the solution of (\ref{Solution_diff_eqn_amplitude}) can be written as
\begin{eqnarray}\label{Solution_Amplitude}
   a=a_{0} \Big(1+\frac{\tau}{T_{r}}\Big)^{-\frac{2}{r}} ,
\end{eqnarray}
where $T_{r}$ is given by the following equation:
\begin{eqnarray}\label{Form_of_T_r}
   T_{r} = \bigg[\frac{r}{4M_{r}} A E \alpha_{1} \sqrt{\frac{r^{2}a_{0}^{r}B_{r}}{(r+1)(r+2)}}  \mathcal{P} \int \limits_{-\infty}^{+\infty}\int \limits_{-\infty}^{+\infty} [\mbox{sech \textit{X}} ]^{\frac{2}{r}} \frac{\partial [\mbox{sech \textit{X}}' ]^{\frac{2}{r}}}{\partial X'}  \frac{dX dX'}{X-X'}\bigg]^{-1} . 
\end{eqnarray}
The equation (\ref{Solution_Amplitude}) gives the amplitude of the solitary wave solution. This equation shows that the amplitude of solitary wave solution is proportional to $\displaystyle  \Big(1+\frac{\tau}{T_{r}}\Big)^{-\frac{2}{r}}$  for $r=1,2,3$.

\begin{center}
\textbf{5. SUMMARY \& DISCUSSIONS}
\end{center} 

\noindent \textbf{(1)} We have considered a collisionless unmagnetized electron-ion plasma consisting of warm adiabatic ions and two distinct populations of electrons at different temperatures - a cooler one is isothermally distributed and follows Boltzmann - Maxwell distribution whereas the hotter one is nonthermally distributed and obeys the distribution function of Cairns et al. [18].

\noindent \textbf{(2)} Considering the Vlasov - Poisson model for two different electron species and the fluid model for ions, we have derived a KdV like evolution equation including the effect of electron Landau damping. We have studied the propagation of weakly nonlinear and weakly dispersive IA waves using this KdV like evolution equation.

\noindent \textbf{(3)} We have seen that the coefficient of the nonlinear term of the KdV like evolution equation vanishes along different family of curves in different parameter planes, viz., $\sigma_{sc}$ - $n_{sc}$, $\beta_{e}$ - $\sigma_{sc}$, $\beta_{e}$ - $n_{sc}$. In this situation, to describe the nonlinear behaviour of  IA waves, we have derived a MKdV like evolution equation including the effect of electron Landau damping having nonlinear term $\displaystyle \big(\phi^{(1)}\big)^{2}\frac{\partial \phi^{(1)}}{\partial \xi}$ but the term responsible for the electron Landau damping remains the same in both KdV and MKdV like evolution equations.

\noindent \textbf{(4)} Again, we have seen that the coefficients of the nonlinear terms of both KdV and MKdV like evolution equations simultaneously vanish along a family of curves for different values of $\sigma$. In this situation, for the first time, we have derived a FMKdV like evolution equation including the effect of electron Landau damping and this equation efficiently describes the nonlinear behaviour of  IA waves. We have found that the nonlinear term of FMKdV like evolution equation is of the form  $\displaystyle \big(\phi^{(1)}\big)^{3}\frac{\partial \phi^{(1)}}{\partial \xi}$ but the term responsible for the electron Landau damping remains same in all KdV, MKdV and FMKdV like evolution equations.

\noindent \textbf{(5)} The evolution equations can be written in a more compact form by considering the nonlinear term of the form $\displaystyle \big(\phi^{(1)}\big)^{r}\frac{\partial \phi^{(1)}}{\partial \xi}$ for $r=1,2,3$. For $r=1,2$ and $3$, we respectively get KdV, MKdV and FMKdV like evolution equations. Using the multiple time scale analysis [45,48,49] with respect to the small parameter $\alpha_{1}$, we have generalised the method of Ott and Sudan [45] to solve the evolution equation (\ref{General_KdV_equation}).

\noindent \textbf{(6)} The solitary wave solution of the evolution equation (\ref{General_KdV_equation}) can be simplified as
\begin{equation}\label{Solution_phi_0_final}
  \phi^{(1)} = a ~  \mbox{sech}^{\frac{2}{r}} X ,
\end{equation}
where
\begin{equation}\label{Expression_of_X_final}
  X = \sqrt{\frac{r^{2}a^{r}B_{r}}{(r+1)(r+2)}} \Big[\xi-\frac{2AB_{r}}{(r+1)(r+2)} \int_{0}^{\tau} a^{r} d\tau\Big] ,
\end{equation}
$a$ is defined by the equation (\ref{Solution_Amplitude}). 

\noindent \textbf{(7)} For the first time, we have found the solitary wave solution of FMKdV like evolution equation and we have seen that the amplitude of solitary wave solution of FMKdV like evolution equation is proportional to $\displaystyle  \Big(1+\frac{\tau}{T_{3}}\Big)^{-\frac{2}{3}}$, where 
$T_{3}$ is given by the equation (\ref{Form_of_T_r}) for $r=3$.

\begin{figure}
\begin{center}
\includegraphics{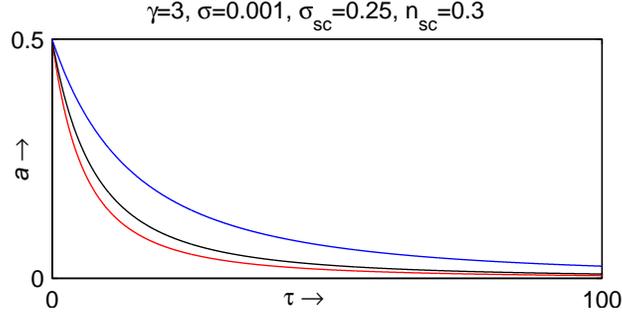}
  \caption{\label{amplitude_vs_tau_KdV_diff_beta_e} The amplitude ($a$) of the KdV soliton is plotted against $\tau$ for different values of $\beta_{e}$. Red, black and blue curves correspond to $\beta_{e}=0$, $\beta_{e}=0.4$ and $\beta_{e}=0.57$ respectively. This figure shows that the amplitude of the KdV soliton decreases with increasing time $\tau$ for any fixed value of $\beta_{e}$ whereas for any fixed value of $\tau$, the amplitude of the KdV soliton increases with the increasing nonthermal parameter $\beta_{e}$.}
\end{center}
\end{figure}

\noindent \textbf{(8)} For $r=1$, the amplitude $a$ of the KdV soliton is plotted against $\tau$ in figure \ref{amplitude_vs_tau_KdV_diff_beta_e} for $\gamma=3$, $\sigma=0.001$, $\sigma_{sc}=0.25$ and $n_{sc}=0.3$ and for different values of $\beta_{e}$. Here red, black and blue curves correspond to $\beta_{e}=0$, $\beta_{e}=0.4$ and $\beta_{e}=0.57$ respectively. From this figure we see that the amplitude $a$ of the KdV soliton increases with increasing $\beta_{e}$ for any fixed $\tau$. This figure also shows that the amplitude decreases with time.   

\begin{figure}
\begin{center}
\includegraphics{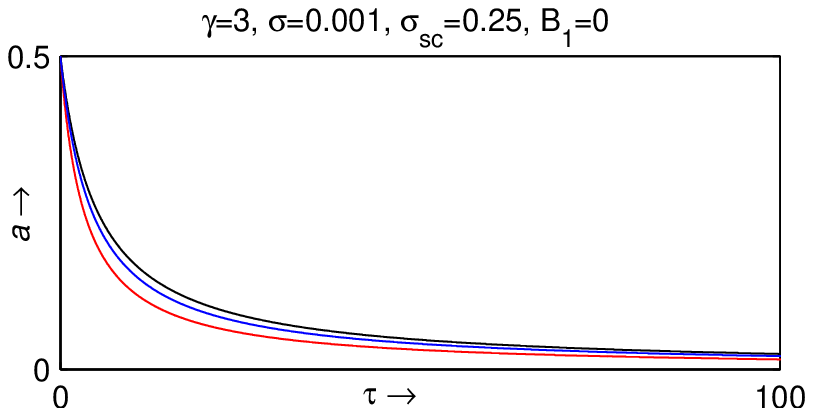}
  \caption{\label{amplitude_vs_tau_MKdV_when_B1_eq_zero} The amplitude ($a$) of the MKdV soliton is plotted against $\tau$ for different values of $\beta_{e}$ when $B_{1}=0$. Red, black and blue curves correspond to $\beta_{e}=0$, $\beta_{e}=0.45$ and $\beta_{e}=0.57$ respectively. This figure shows that the amplitude of the MKdV soliton decreases with increasing time $\tau$ for any fixed value of $\beta_{e}$.}
\end{center}
\end{figure}

\noindent \textbf{(9)} For $r=2$, the amplitude $a$ of the MKdV soliton is plotted against $\tau$ in figure \ref{amplitude_vs_tau_MKdV_when_B1_eq_zero} when $B_{1}=0$ for $\gamma=3$, $\sigma=0.001$ and $\sigma_{sc}=0.25$, and for different values of $\beta_{e}$. Here red, black and blue curves correspond to $\beta_{e}=0$, $\beta_{e}=0.45$ and $\beta_{e}=0.57$ respectively. This figure shows that the amplitude decreases with time.   

\begin{figure}
\begin{center}
\includegraphics{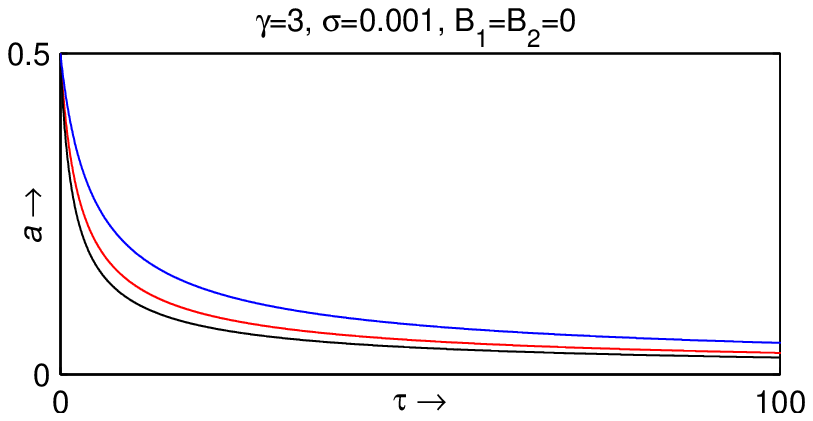}
  \caption{\label{amplitude_vs_tau_FMKdV_when_B1_B2_eq_zero_diff_beta_e} The amplitude ($a$) of the FMKdV soliton is plotted against $\tau$ for different values of $\beta_{e}$ when $B_{1}=B_{2}=0$. Red, black and blue curves correspond to $\beta_{e}=0$, $\beta_{e}=0.352$ and $\beta_{e}=0.42$ respectively. This figure shows that the amplitude of the FMKdV soliton decreases with increasing time $\tau$ for any fixed value of $\beta_{e}$.}
\end{center}
\end{figure}

\noindent \textbf{(10)} For $r=3$, the amplitude $a$ of the FMKdV soliton is plotted against $\tau$ in figure \ref{amplitude_vs_tau_FMKdV_when_B1_B2_eq_zero_diff_beta_e} when $B_{1}=B_{2}=0$ for $\gamma=3$ and $\sigma=0.001$, and for different values of $\beta_{e}$. Red, black and blue curves correspond to $\beta_{e}=0$, $\beta_{e}=0.352$ and $\beta_{e}=0.42$ respectively. This figure shows that the amplitude decreases with time.   

\noindent \textbf{(11)} Therefore, from figure \ref{amplitude_vs_tau_KdV_diff_beta_e}  - figure \ref{amplitude_vs_tau_FMKdV_when_B1_B2_eq_zero_diff_beta_e}, we can conclude that the amplitude of the IA soliton decreases with time $\tau$ for all $r=1,2,3$ if the effect of electron Landau damping is taken into account.  

\noindent \textbf{APPENDIX - A}  Coefficients of the equation (\ref{Vlasov_f_ce7_modified}):
\begin{equation}\label{J_c0}
   J_{c0} = \int_{-\infty}^{+\infty} \frac{\partial j_{c0}}{\partial v_{||}^{2}}dv_{||}~,~Z_{c0} = \frac{\partial f_{c0}}{\partial v_{||}^{2}}\bigg|_{v_{||}=0},
\end{equation}
where $J$ = $F$, $G$, $H$, $K$ for $j$ = $f$, $g$, $h$, $k$ respectively.

\noindent \textbf{APPENDIX - B} Coefficients of the equation (\ref{Vlasov_f_se7}):
\begin{equation}\label{J_s0}
   J_{s0} = \int_{-\infty}^{+\infty} \frac{\partial j_{s0}}{\partial v_{||}^{2}}dv_{||}~,~Z_{s0} = \frac{\partial f_{s0}}{\partial v_{||}^{2}}\bigg|_{v_{||}=0},
\end{equation}
where $J$ = $F$, $G$, $H$, $K$ for $j$ = $f$, $g$, $h$, $k$ respectively.

\noindent \textbf{APPENDIX - C}  Method of finding $I_{r}$ associated with the equations (\ref{Solution_expression_W_r}) and (\ref{Solution_expression_I_r}):
\begin{equation}\label{Cauchy_principal_value}
   I_{r} = \mathcal{P} \int \limits_{-\infty}^{+\infty}\int \limits_{-\infty}^{+\infty} [\mbox{sech \textit{X}} ]^{\frac{2}{r}} \frac{\partial [\mbox{sech \textit{X}}' ]^{\frac{2}{r}}}{\partial X'}  \frac{dX dX'}{X-X'}   .
\end{equation}
Now $I_{r}$ can be written as 
\begin{equation}\label{form_of_I}
   I_{r} = - \int_{-\infty}^{\infty} \frac{\partial [\mbox{sech \textit{z}} ]^{\frac{2}{r}}}{\partial z} I_{1r} dz ,
\end{equation}
where $X=z'$, $X'=z $ and
\begin{equation}\label{form_of_II}
   I_{1r} = \mathcal{P} \int_{-\infty}^{\infty}  \frac{[\mbox{sech \textit{z}}' ]^{\frac{2}{r}} }{z-z'} dz' .
\end{equation}
Using the following known result 
\begin{equation}\label{Fourier_representation_of_Integral}
   \int_{-\infty}^{0} e^{is(z-z')} ds = \pi \delta (z-z') - i \mathcal{P} \frac{1}{z-z'} , 
\end{equation}
form the equation (\ref{Fourier_representation_of_Integral}), we get
\begin{equation}\label{Fourier_representation_of_Integral_P}
   \mathcal{P} \frac{1}{z-z'} = \frac{1}{2i} \int_{-\infty}^{\infty} \frac{s}{|s|} e^{is(z-z')} ds  .
\end{equation}
Using (\ref{Fourier_representation_of_Integral_P}), equation (\ref{form_of_II}) can be written as
\begin{equation}\label{form_of_I_1}
   I_{1r} = \frac{1}{2i} \int_{-\infty}^{\infty} \frac{s}{|s|} F(s) e^{isz} ds , 
\end{equation}
where
\begin{equation}\label{the_inegral_of_F_k}
   F(s) = \int_{-\infty}^{\infty} [\mbox{sech \textit{z}} ]^{\frac{2}{r}} e^{-isz} dz .
\end{equation}
Therefore, the equation (\ref{form_of_I}) can be written as
\begin{equation}\label{the_total_integral_of_I}
   I_{r} = \int_{0}^{\infty} s [F(s)]^{2}  ds .
\end{equation}
\newpage
\begin{center}
\textbf{REFERENCES}
\end{center}
\begin{enumerate}
\item W. D. Jones, A. Lee, S. M. Gleman,  and H. J. Doucet, Phys. Rev. Lett. \textbf{35}, 1349 (1975).

\item 
N. Hershkowitz, Space Sci. Rev. \textbf{41}, 351 (1985).

\item 
Y. Nishida and T. Nagasawa, Phys. Fluids \textbf{29}, 345 (1986).

\item 
G. Hairapetian and R. L. Stenzel, Phys. Rev. Lett. \textbf{65}, 175 (1990).

\item 
W. C. Feldman, J. R. Asbridge, S. J. Bame, M. D. Montgomery,   and S. P. Gary, J. Geophys. Res. \textbf{80}, 4181 (1975).

\item 
M. Temerin, K. Cerny, W. Lotko,  and F. S. Mozer, Phys. Rev. Lett. \textbf{48}, 1175 (1982).

\item 
P. O. Dovner, A. I. Eriksson, R. Bostr{\"o}m,  and B. Holback, Geophys. Res. Lett. \textbf{21}, 1827 (1994).

\item 
R. E. Ergun, C. W. Carlson, J. P. McFadden, F. S. Mozer, G. T. Delory, W. Peria, C. C. Chaston,  M. Temerin,  R. Elphic,  R. Strangeway,  et al., Geophys. Res. Lett. \textbf{25}, 2025 (1998).

\item 
R. E. Ergun, C. W. Carlson, J. P. McFadden, F. S. Mozer, G. T. Delory, W. Peria, C. C. Chaston,  M. Temerin,  R. Elphic,  R. Strangeway,  et al., Geophys. Res. Lett. \textbf{25}, 2061 (1998).

\item 
G. T. Delory, R. E. Ergun, C. W. Carlson, L. Muschietti, C. C. Chaston, W. Peria, J. P. McFadden,  and R. Strangeway, Geophys. Res. Lett. \textbf{25}, 2069 (1998).

\item 
R. Pottelette, R. E. Ergun, R. A. Treumann, M. Berthomier, C. W. Carlson, J. P. McFadden,  and I. Roth, Geophys. Res. Lett. \textbf{26}, 2629 (1999).

\item 
J. P. McFadden, C. W. Carlson, R. E. Ergun, F. S. Mozer, L. Muschietti, I. Roth,  and E. Moebius, J. Geophys. Res. \textbf{108}, 8018 (2003).

\item 
R. Bostr{\"o}m, G. Gustafsson, B. Holback, G. Holmgren, H. Koskinen,  and P. Kintner, Phys. Rev. Lett. \textbf{61}, 82 (1988).

\item 
R. Bostr{\"o}m, IEEE Trans. Plasma Sci. \textbf{20}, 756 (1992).

\item 
H. Matsumoto, H. Kojima, T. Miyatake, Y. Omura, M. Okada, I. Nagano,  and M. Tsutsui, Geophys. Res. Lett. \textbf{21}, 2915 (1994).

\item 
J. R. Franz, P. M. Kintner,  and J. S. Pickett, Geophys. Res. Lett. \textbf{25}, 1277 (1998).

\item 
C. A. Cattell, J. Dombeck, J. R. Wygant, M. K. Hudson, F. S. Mozer, M. A. Temerin, W. K. Peterson, C. A. Kletzing, C. T. Russell,  and R. F. Pfaff, Geophys. Res. Lett. \textbf{26}, 425 (1999).

\item 
R. A. Cairns, A. A. Mamum, R. Bingham, R. Bostr{\"o}m, R. O. Dendy, C. M. C. Nairn, and P. K. Shukla, Geophys. Res. Lett. \textbf{22}, 2709 (1995).

\item 
B. N. Goswami and B. Buti, Phys. Lett. A \textbf{57}, 149 (1976).

\item 
K. Nishihara and M. Tajiri, J. Phys. Soc. Jpn \textbf{50}, 4047 (1981).

\item 
R. Bharuthram and P. K. Shukla, Phys. Fluids \textbf{29}, 3214 (1986).

\item 
L. L. Yadav, R. S. Tiwari, K. P. Maheshwari,  and S. R. Sharma, Phys. Rev. E \textbf{52}, 3045 (1995).

\item 
S. G. Tagare, Phys. Plasmas \textbf{7}, 883 (2000).

\item 
S. Islam, A. Bandyopadhyay,  and K. P. Das, J. Plasma Phys. \textbf{74}, 765 (2008).

\item 
A. E. Dubinov, and M. A. Sazonkin, Plasma Phys. Rep. \textbf{35}, 14 (2009).

\item 
O. R. Rufai, R. Bharuthram, S. V. Singh,  and G. S. Lakhina, Phys. Plasmas \textbf{21}, 082304 (2014).

\item 
S. V. Singh and G. S. Lakhina, Commun. Nonlinear Sci. Numer. Simul. \textbf{23}, 274 (2015).

\item 
D. N. Gao, J. Zhang, Y. Yang, and W. S. Duan, Plasma Phys. Rep. \textbf{43}, 833 (2017).

\item 
S. Dalui, A. Bandyopadhyay,  and K. P. Das, Phys. Plasmas \textbf{24}, 042305 (2017).

\item 
S. Dalui, A. Bandyopadhyay,  and K. P. Das, Phys. Plasmas \textbf{24}, 102310 (2017).

\item 
A. Vlasov, J. Phys. (U.S.S.R.) \textbf{9}, 25 (1945).

\item 
L. D. Landau, J. Phys. (U.S.S.R.) \textbf{10}, 25 (1946).

\item 
D. Bohm and E. P. Gross, Phys. Rev. \textbf{75}, 1851 (1949).

\item 
N. G. Van Kampen, Physica \textbf{21}, 949 (1955).

\item 
P. L. Auer, Phys. Rev. Lett. \textbf{1}, 411 (1958).

\item 
K. M. Case, Annals Phys. \textbf{7}, 349 (1959).

\item 
O. Penrose, Phys. Fluids \textbf{3}, 258 (1960).

\item 
J. Dawson, Phys. Fluids \textbf{4}, 869 (1961).

\item 
T. O'Neil, Phys. Fluids \textbf{8}, 2255 (1965).

\item 
R. Z. Sagdeev and A. A. Galeev, \textit{Nonlinear plasma theory} (Edited by T. M. O'Neil, W. A. Benjamin: New York, 1969) pp. 37-43.

\item 
J. D. Meiss and P. J. Morrison, Phys. Fluids \textbf{26}, 983 (1983).

\item 
V. P. Silin, Plasma Phys. Rep. \textbf{39}, 1055 (2013).

\item 
J. H. Malmberg and C. B. Wharton, Phys. Rev. Lett. \textbf{13}, 184 (1964).

\item 
N. I. Grishanov and N. A. Azarenkov, Plasma Phys. Rep. \textbf{39}, 947 (2013).

\item 
E. Ott and R. N. Sudan, Phys. Fluids \textbf{12}, 2388 (1969).

\item 
J. W. VanDam and T. Taniuti, J. Phys. Soc. Jpn. \textbf{35}, 897 (1973).

\item 
M. Tajiri and K. Nishihara, J. Phys. Soc. Jpn. \textbf{54}, 572 (1985).

\item 
A. Bandyopadhyay and K. P. Das, Phys. Plasmas \textbf{9}, 465 (2002).

\item 
A. Bandyopadhyay and K. P. Das, Phys. Plasmas \textbf{9}, 3333 (2002).

\item 
S. Ghosh and R. Bharuthram, Astrophys. Space Sci. \textbf{331}, 163 (2011).

\item 
A. P. Misra and A. Barman, Phys. Plasmas \textbf{22}, 073708 (2015).

\item 
A. Barman and A. P. Misra, Phys. Plasmas \textbf{24}, 052116 (2017).

\item 
Y. Saitou and Y. Nakamura, Phys. Lett. A \textbf{343}, 397 (2005).

\item 
Y. Ghai, N. S. Saini,  and B. Eliasson, Phys. Plasmas \textbf{25}, 013704 (2018).

\item 
J. Weiland, Y. H. Ichikawa,  and H. Wilhelmsson, Phys. Scr. \textbf{17}, 517 (1978).

\item 
S. Wolfram, \textit{The Mathematica {\textregistered} Book, Version 4} (Cambridge university press Cambridge, 1999).
\end{enumerate}
\newpage

\end{document}